\documentclass[]{elsarticle}

\usepackage[normalem]{ulem}

\usepackage{lineno,hyperref}
\usepackage{amsmath,amssymb}
\usepackage{caption} 
\usepackage{multirow}
\usepackage{xcolor}

\biboptions{authoryear}

\newcommand{\x}{\mathbf{x}}
\newcommand{\y}{\mathbf{y}}
\newcommand{\X}{\mathbf{X}}
\newcommand{\rr}{\mathbf{r}}
\newcommand{\ttt}{\mathbf{t}}
\newcommand{\RR}{\mathbf{R}}

\newcommand{\R}{\mathbb{R}}
\newcommand{\M}{\mathbb{M}}

\begin{document}

\begin{frontmatter} 

\title{Fitting three-dimensional Laguerre tessellations by hierarchical marked point process models}


\author[mymainaddress]{Filip Seitl\corref{mycorrespondingauthor}}
\cortext[mycorrespondingauthor]{Corresponding author}
\ead{seitl@karlin.mff.cuni.cz}

\author[mysecondaryaddress]{Jesper M{\o}ller}

\author[mymainaddress]{Viktor Bene\v{s}}

\address[mymainaddress]{Department of Probability and Mathematical Statistics, Faculty of Mathematics and Physics, Charles University, Sokolovsk\'{a} 83, 186 75, Praha 8, Czech Republic}
\address[mysecondaryaddress]{Department of Mathematical Sciences, Aalborg University, Skjernvej 4A, DK-9220~Aalborg~\O, Denmark}

\begin{abstract}
{We present a general statistical methodology for analysing a Laguerre tessellation data set viewed  
as a realization of a marked point process model.
In the first step, for the point,
we use a nested sequence of multiscale processes which constitute a flexible parametric class of pairwise interaction point process models. In the second step, for the marks/radii conditioned on the points,
we consider various exponential family models where the canonical sufficient statistic is based on tessellation characteristics. For each step, parameter estimation based on maximum pseudolikelihood methods is tractable}. {For model selection, we consider maximized log pseudolikelihood functions for models of the radii conditioned on the points.} Model checking is performed using global envelopes and corresponding tests in {both steps and moreover} by comparing observed and simulated tessellation characteristics in the second step. We apply our methodology for a 3D Laguerre tessellation data set representing the microstructure of a polycrystalline metallic material, where simulations under a fitted model may substitute expensive laboratory experiments.
\end{abstract}


\end{frontmatter}


\section{Introduction}\label{intro}

This paper develops new statistical methodology for analysing Laguerre tessellation data sets, 
where we illustrate how the methodology applies for an example of a 3D Laguerre 
tessellation data set representing a polycrystalline material. 




\subsection{Some background and motivation}\label{s:moti}


{In brief, a 3D Laguerre tessellation (also called a power diagram or a generalised Voronoi tessellation) is a flexible way of modelling a subdivision of 3D space (see Section~\ref{intro.1} for the details). In practice the tessellation is described by a data set $(\x_n,\rr_n)$ where $\x_n=\{x_1,\ldots,x_n\}\subset\R^3$ is a finite point pattern and 
$\rr_n=(r_1,\ldots,r_n)\in(0,\infty)^n$ is an associated vector of positive numbers called marks, and 
we identify $(\x_n,\rr_n)$ by the marked point pattern 
$\{(x_1,r_1),\ldots,(x_n,r_n)\}$
 (see again Section~\ref{intro.1} for the details).}
%
Such data sets appear in various contexts of material science. For example, 
in 3D X-ray diffraction microscopy 
{of polycrystalline materials the volumes and centroids of grains are obtained, and representation of such measurements }
by a 3D Laguerre tessellation can be produced by various optimization methods \citep{Lyckegaard,spettl2016,QueyR,kuhn2020}. It is 
then 
useful to develop stochastic models for the tessellation since simulations can to some extent substitute expensive laboratory experiments. However, only a few papers  \citep[including][]{spettl2015,seitl2020} have been dealing with statistical methodology in the context of polycrystalline materials. 

Inference for statistical models of Laguerre tessellation data sets is a difficult task unless $(\x_n,\rr_n)$ follows a simple model such as a (marked) Poisson process which in practice is rarely a reasonable assumption. Often the points in $\x_n$ exhibit regularity and different marked point process models for $(\x_n,\rr_n)$ have been suggested: simple random models for packings of hard balls \citep{chiu2013} obtained by iterative procedures such as random sequential adsorption \citep[][with applications to foam structures]{lautensack2008a} or variants of collective-rearrangement algorithms \citep[][with application to a polycrystalline microstructure]{spettl2015}; and 
inspired by the work in \cite{dereudre2011} on Gibbsian models for 2D Voronoi tessellations, \cite{seitl2020} introduced Gibbsian models for random 3D Laguerre tessellations. These Gibbs models were first thought to be tempting to use because they may incorporate properties of the Laguerre tessellation and interaction between its cells. However, the models are 
complicated to use for statistical inference 
and they are time-consuming to simulate. 

\subsection{Our contribution}\label{s:intro.2}

In this paper we introduce hierarchical statistical models consisting of first a~parametric Gibbs point process model for $\x_n$ and second a parame\-tric model for $\rr_n$ conditioned on $\x_n$. Specifically, we use in the first case a nested sequence of flexible pairwise interaction points processes called multiscale processes \citep{penttinen1984} and in the second case various exponential family models where the canonical sufficient statistic is based on tessellation characteristics such as surface area or volume of cells or absolute difference in volumes of neighbouring cells. Apart from reducing the dimension from 4 (when viewing $(\x_n,\rr_n)$ as a~4-dimensional point pattern) to 3 (when considering $\x_n$), an advantage is that we specify two much simpler cases of models with parameters which do not depend on each other. Hence we can separate between how to simulate and estimate unknown parameters for $\x_n$  and $\rr_n\mid \x_n$, respectively. The 
  parameters are simply estimated by maximum pseudolikelihood methods and well-known MCMC algorithms are used for si\-mulations. Thereby estimation and simulation become much faster than in \cite{seitl2020} when fitting specific models given in Section~\ref{sec:hm} and used for analysing the data from Section~\ref{sec:data} 
in Section~\ref{sec:modsel}.  

A further advantage is that the model construction makes it possible to develop a rather straightforward model selection procedure: 
For $\x_n$, the procedure starts with the simplest case of a Poisson process and continues with constructing more and more complex multiscale processes until a satisfactory fit is obtained when considering global envelopes and tests \citep{mm2017} based on various functional summary statistics. 
For $\rr_n$ conditioned on $\x_n$, more and more complex exponential models are developed, where we demonstrate how to compare fitted models of the same dimension by considering maximized log pseudolikelihood functions. Further, we evaluate selected fitted models by comparing moment properties of tessellation characteristics under simulations from the model with empirical moments, by considering plots of global envelopes, and by evaluating values of global envelope tests.
 This comparison is not only done by looking at those tessellation characteristics used for specifying the canonical sufficient statistic of the exponential model but also for various other tessellation characteristics.
To the best of our knowledge, this is the first time that such a model selection procedure has been used when analysing  
polycrystalline materials, cf.\ \cite{sed2018} and the references therein. 

For the computations of Laguerre tessellations, we use the \texttt{C++} library \texttt{Voro++} \citep{rycroft2009}. 
This open source library enables computation of single Laguerre cells which is an advantage for the estimation and MCMC simulation procedures we use. 
For the calculation of
functional summary statistics (specifically $\hat L$, $\hat F$ and $\hat G$), we use
the \texttt{R}-package \texttt{spatstat} \citep{baddeley2015}, and for
 global envelopes and corresponding $p$-values we use the \texttt{R}-package \texttt{GET} \citep{myllymaki2019}.
We implemented a code for pseudolikelihood estimation and the MCMC algorithms used for simulation of the point process models and the models for the radii given the points. This code is available at \url{https://github.com/VigoFierry/Lag_mod}.
 

The paper is organised as follows. 
Section~\ref{intro.1} provides details on Laguerre tessellation data sets. Section~\ref{sec:data} describes a particular data set which in later sections is used to illustrate our statistical methodology. Section~\ref{sec:hm} specifies our hierarchical model construction and discusses how we used maximum pseudolikelihood estimation and simulated under fitted models for $\x_n$ and $\rr_n\mid\x_n$.
Section~\ref{sec:modsel} details our model selection procedure and applies it on the Laguerre tessellation data set from Section~\ref{sec:data}. 
Finally, Section~\ref{sec:con} contains concluding remarks. 
The paper is accompanied by supplementary material available in appendices A-E in this preprint.
 


\section{Laguerre tessellation data sets}\label{intro.1}

A Laguerre tessellation is defined as follows. Consider a marked point pattern $\{(x_j,r_j); j\in J\}\subset\R^3\times(0,\infty)$ where 
each $x_j \in \R^3$ is a spatial location with an associated mark $r_j>0$ and the index set $J$ is finite or countable. Define the corresponding point configuration $\x=\{x_j; j\in J\}$ and the associated (possibly infinite) vector of marks $\rr=(r_j;j\in J)$ (using some arbitrary ordering of $J$). 
Interpret
$(x_j,r_j)$ as a closed ball 
with center $x_j$ and radius $r_j$, and define the power distance from a point $y\in\R^3$ to this ball by
\begin{equation*}
    \label{e:dist}
    \rho(y,(x_j,r_j)) = \lVert y-x_j \rVert^2 - r_j^2
\end{equation*}
where $\|\cdot\|$ denotes Euclidean distance. Assuming that $\min_{j\in J} \rho(y,(x_j,r_j))$ exists for every $y\in\R^3$, the Laguerre cell with generator $(x_j,r_j)$ is
 \begin{equation*}\label{e:def}
C(x_j,r_j\mid \x,\rr) = \bigcap_{k\in J}\{ y \in \R^3 \mid \rho(y,(x_j,r_j)) \leq \rho(y,(x_k,r_k))\}
\end{equation*}
and
 the Laguerre tessellation generated by $(\x,\rr)$ is
the collection of all non-empty cells $C(x_j,r_j\mid \x,\rr)$. 

Notice that $\{(x_j,r_j); j\in J\}$ and $(\x,\rr)$ are in one-to-one correspondence. We use this identification without any mentioning in the following. But it should be kept in mind when we abuse terminology by calling $(\x,\rr)$ a marked point pattern, and when we abuse notation and write e.g.\ $((x,y,z),r)\in (\x,\rr)$ (which means that $((x,y,z),r)= (x_j,r_j)$ for some $j\in J$).

We refer to \cite{lautensack2008} and the references therein for detailed studies of deterministic and random Laguerre tessellations. The reason why Laguerre tessellations are so 
{useful for modelling purposes} may be explained by a fundamental result due to \cite{aurenhammer1987a} and extended by \cite{lautensack2008} to infinite cases: any 3D (or higher-dimensional) normal tessellation 
with convex cells is a Laguerre tessellation.

{In Section~\ref{sec:data} and later sections we consider the following setting.} 
Let $W=[0,a]\times[0,b]\times[0,c]$ be a rectangular parallelepiped and $(\x,\rr)$
 a finite marked point pattern  such that $\x\subset W$. Define
 a periodic extension of $(\x,\rr)$ which is in accordance to $W$, i.e., the infinite marked point pattern
\begin{equation}\label{e:jm0}
(\x^*,\rr^*) = \bigcup_{((x,y,z),r) \in (\x,\rr)}\bigcup_{(i,j,k)\in {\mathbb Z}^3} \{((ia+x,jb+y,kc+z),r)\}
\end{equation}
where $\mathbb{Z}^3$ is the 3D integer lattice. Note that $\x=\x^*\cap W$. Then we consider the Laguerre tessellation generated by $(\x^*,\rr^*)$ and keep only those Laguerre cells $C(x_j,r_j\mid\x^*,\rr^*)$ which are non-empty and have $x_j\in\x$. Let $n$ be the number of such cells and $(\x_n,\rr_n)\subseteq (\x,\rr)$ be the marked point pattern specifying the generators of these cells. Then $(\x_n,\rr_n)$ is the Laguerre tessellation data set used for {our} statistical analysis. 

{This construction of $(\x_n,\rr_n)$ is one way of how Laguerre tessellation data sets may be produced, and we will consider $(\x_n,\rr_n)$ as a realisation of a finite marked point process \citep[for background material on marked point processes, see e.g.][]{chiu2013}. Note that the statistical methodology introduced in this paper can immediately be extended to other settings with a finite marked point pattern data set representing a Laguerre tessellation.}


\section{Application example}\label{sec:data}


The Laguerre tessellation data set used throughout the text to illustrate our statistical methodology is related to a study in \cite{sedmak2016} of a~polycrystalline microstructure of a nickel titanium alloy. 
Using a notation as in Section~\ref{intro.1},  
we deal with a~finite marked point pattern $(\x,\rr)$ 
extracted from a~larger data set collected in \cite{petrich2019} by the so-called cross-entropy method applied to 3D X-ray diffraction measurements. Specifically, $\x$ consists of the 2009 points which are contained in a  
 3D rectangular observation window $W$ of size  
$40\times 40\times 85$ (the units being micrometers), where $W$ is a cut-off from the entire material specimen. 
{Some of the cells in the Laguerre tessellation generated by $(\x,\rr)$ are unbounded and hence not contained in the material specimen. 
To account for this and for computational reasons (i.e., the use of \texttt{Voro++}), we choose to apply a periodic extension thereby obtaining our Laguerre tessellation data set 
$(\x_n,\rr_n)$, cf.\ Section~\ref{intro.1}.
Here $n=1965$, }
so by applying the periodic extension at most 44 cells are `lost'. 

The left panel in Figure~\ref{fig:generators} below depicts $\x_n$ together with $W$ and \ref{ap:1} and \ref{ap:2} show images of the observed Laguerre tessellation together with plots of the projections of $\x_n$ onto the $xy$, $xz$ and $yz$ planes, respectively. None of these indicate
spatial inhomogeneity of $\x_n$.
So considering the periodic extension $\x^*_n$ (defined as in \eqref{e:jm0} but with $(\x,\rr)$ replaced by $(\x_n,\rr_n)$),
we find it reasonable to assume that the
distribution underlying $\x^*$ is invariant under translations in 3D space. Equivalently, we assume 
that the
distribution underlying 
$\x_n$ is invariant under shifts when 
$W$ is wrapped on a 3D torus.
The right panel in Figure~\ref{fig:generators} shows the balls defined by $(\x_n,\rr_n)$. This indicates
that the distribution underlying the radii $\rr_n$ conditioned on $\x_n$ can be assumed to be homogeneous, i.e., we have invariance in this conditional distribution when making shifts of $\x_n$ on the torus. This assumption is further supported by plots in \ref{ap:2} showing the radii distributions corresponding to a subdivision of $W$ into sets of equal size and shape. Moreover, Figure~\ref{fig:radii} below shows a histogram of the radii together with a kernel density estimate and the density of a beta distribution fitted by maximum pseuodolikelihood estimation as discussed later in Sections~\ref{sec:hm} and \ref{sec:modsel}.


\begin{figure}[h!]
\center
\begin{minipage}{0.42\textwidth}
\includegraphics[width=1\linewidth]{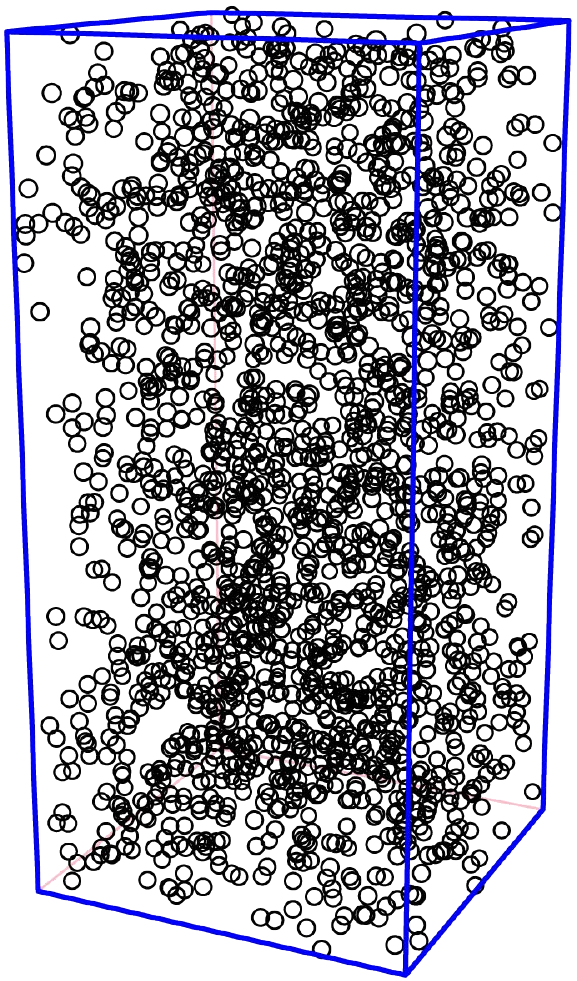}
\end{minipage} 
\begin{minipage}{0.42\textwidth}
\includegraphics[width=1.2\linewidth]{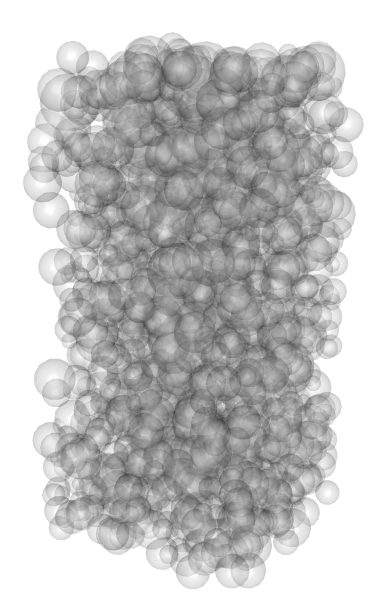}
\end{minipage} 
\caption{The observation window $W$ and the point pattern $\x_n$ 
(left panel) 
and the balls defined by $(\x_n,\rr_n)$ (right panel).}
\label{fig:generators}
\end{figure}


\begin{figure}[h!]
\center
\includegraphics[width=0.5\linewidth]{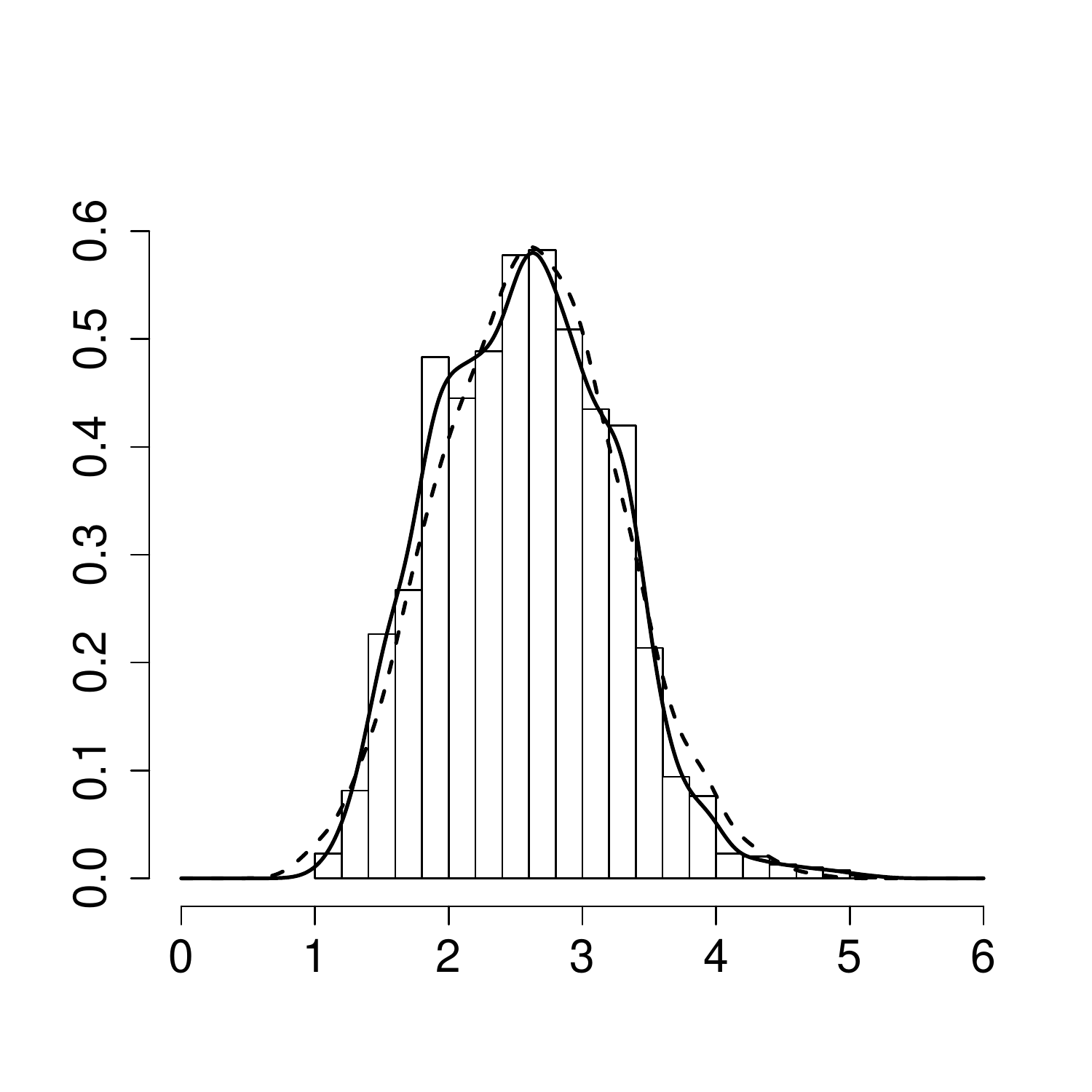}
\caption{Histogram of the radii together with a kernel density estimate (solid line) and a fitted density of a beta distribution (dashed line) obtained by maximum pseudolikelihood estimation.}
\label{fig:radii}
\end{figure}

\begin{figure}[h!]
\center
\includegraphics[width=0.3\linewidth]{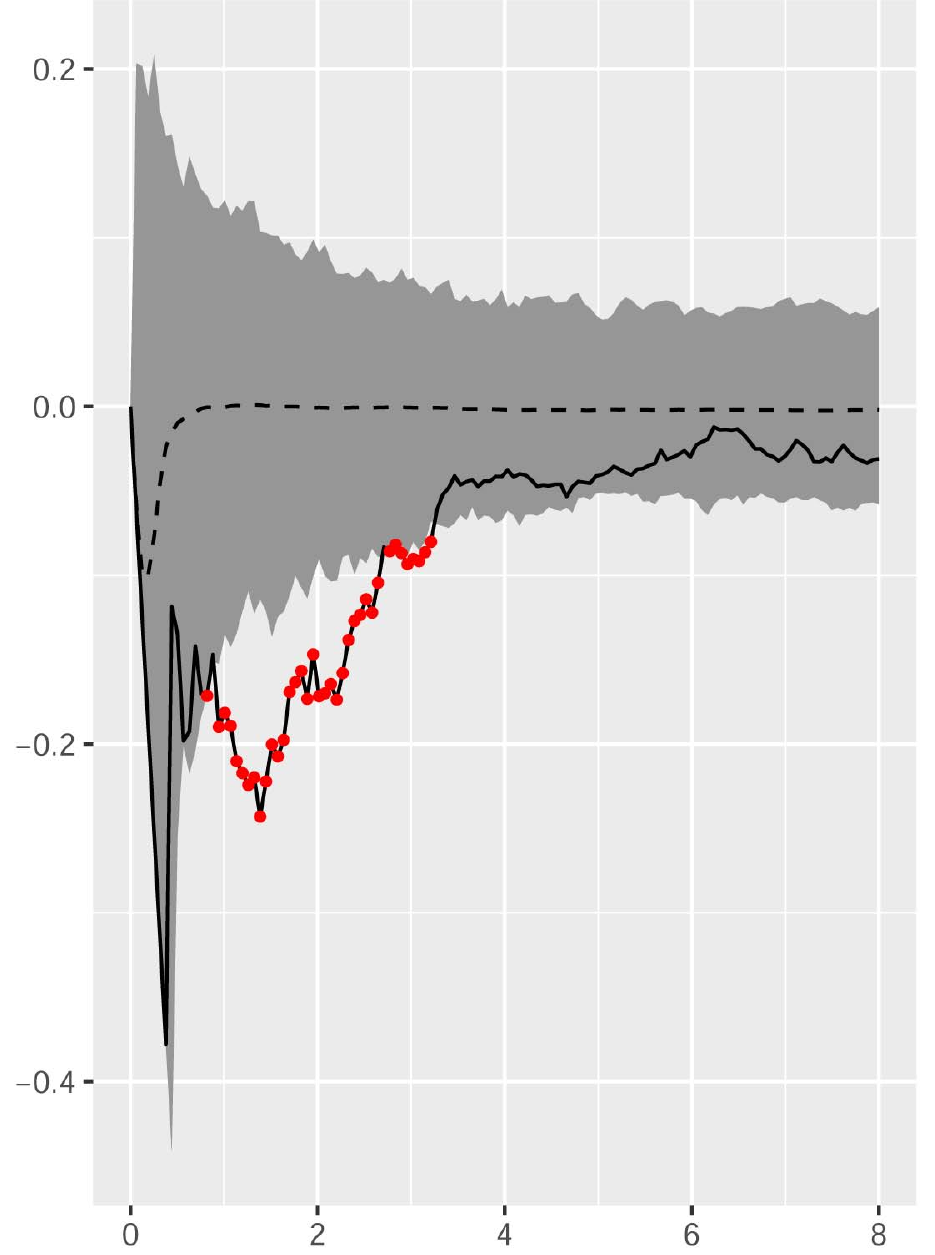}
\includegraphics[width=0.3\linewidth]{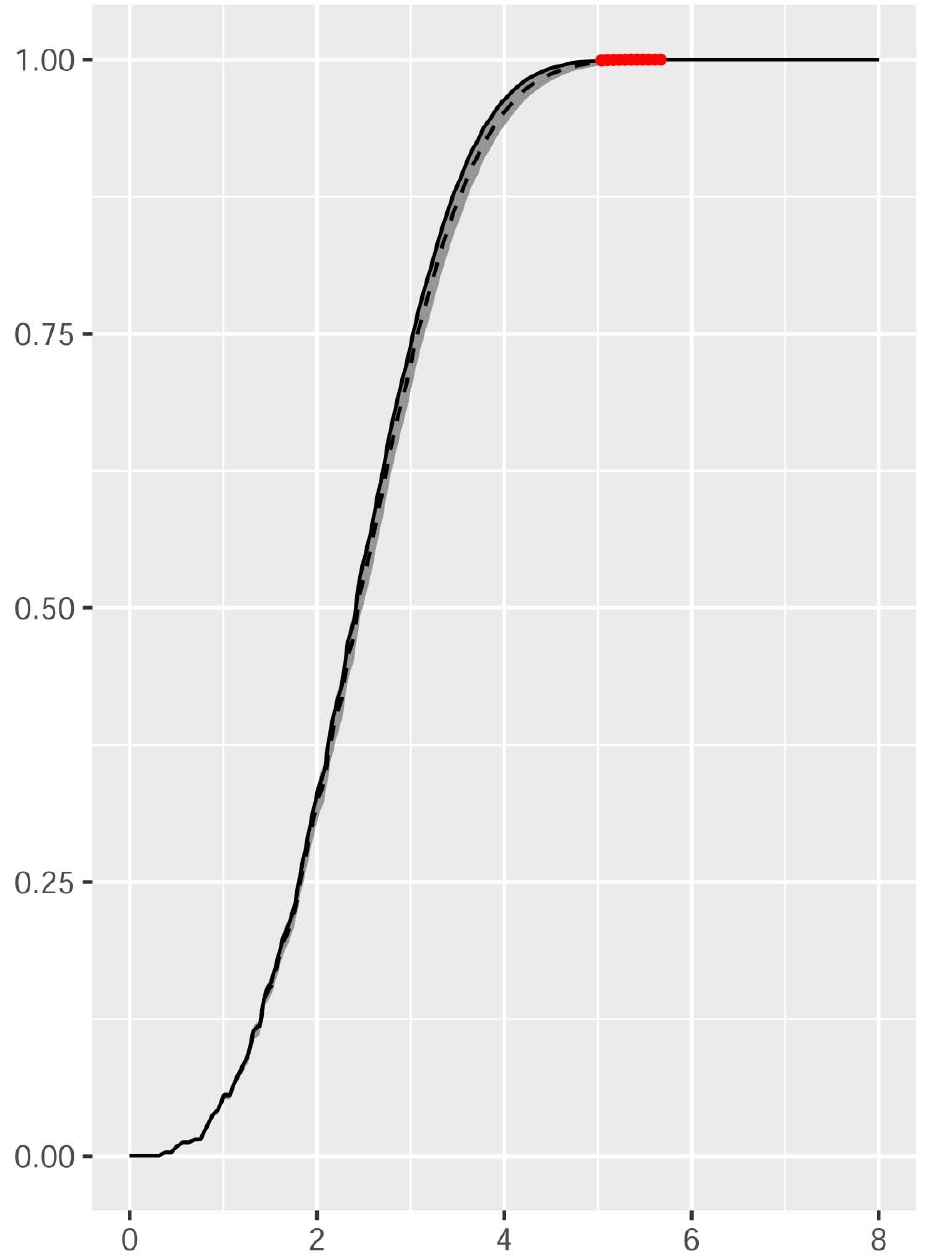}
\includegraphics[width=0.3\linewidth]{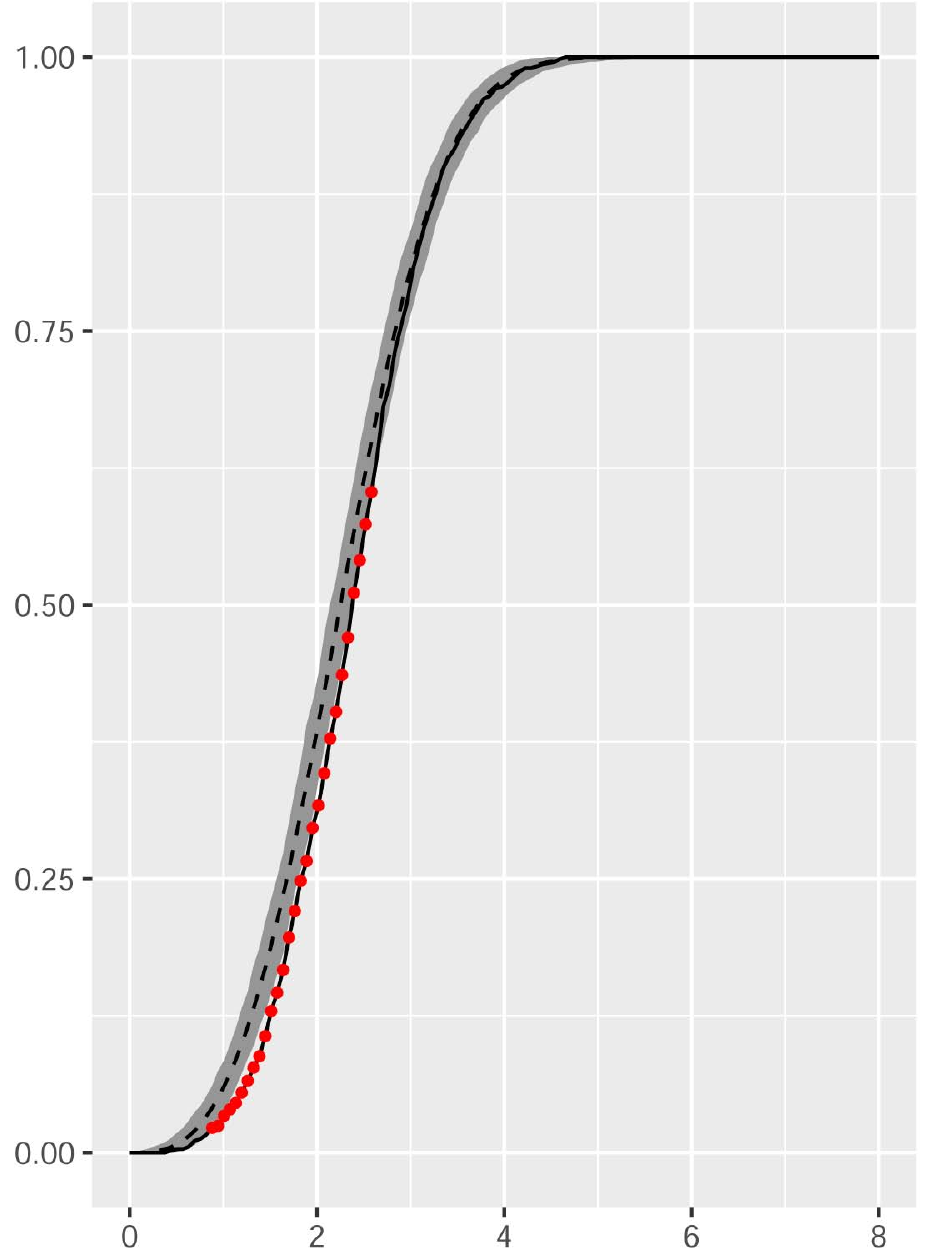}
\caption{From left to right, empirical functional summary statistics $\hat L(t)-t$, $\hat F(t)$ and $\hat G(t)$ (solid lines) and simulated 95\%-global envelope (grey regions) obtained under a fitted homogeneous Poisson process. Dashed lines are averages of the simulated functional summary statistics 
and the dots indicate when the empirical functions are outside the envelope.} 
\label{fig:PP0_GET} 
\end{figure}

Figure~\ref{fig:PP0_GET} shows estimated/empirical functional summary statistics 
$\hat L(t)-t$, $\hat F(t)$ and $\hat G(t)$
for the point pattern $\x_n$ where we used 
the \texttt{R}-package \texttt{spatstat} \citep{baddeley2015} for the calculations.
For definitions and interpretations of these empirical functions using edge correction factors, see e.g.\ \cite{baddeley2015} 
 (for 
$\hat L$ we used Ripley's isotropic edge correction factor and for
$\hat F$ and $\hat G$ we used Kaplan-Meier edge correction factors). Figure~\ref{fig:PP0_GET} also shows a concatenated 
95\%-global envelope/confidence region (the grey region) under a homogeneous Poisson process (with the intensity estimated by $n$ divided by the volume of $W$), meaning that under this Poisson process all three empirical functions are expected to be within the envelope with probability 0.95. The envelopes were obtained using the \texttt{R}-package \texttt{GET} \citep{myllymaki2019} with 1999 simulations of the Poisson process (increasing this to 9999 simulations did not change the results). The plots show that the empirical functions are often outside the envelope, and the corresponding $p$-value obtained by the global area rank envelope test \citep{mm2017}
is below 0.1\% \citep[there are several possible choices of making an envelope test, but the area rank envelope test is recommended in][]{myllymaki2020}. The way the functions based on the data differ from the envelope indicates regularity  in the point pattern $\x_n$. 

\begin{figure}[h!]
\includegraphics[width=0.24\linewidth]{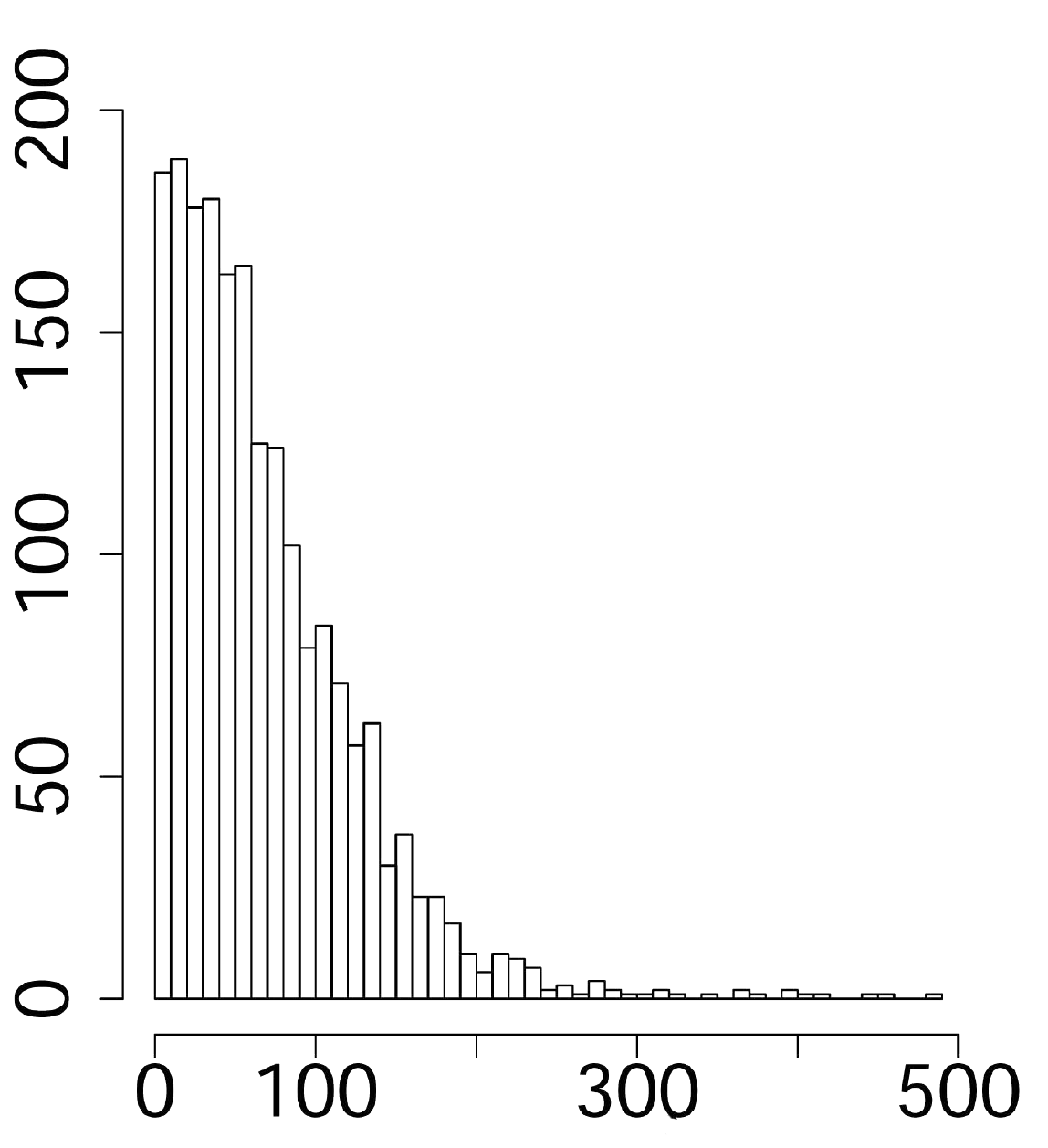}
\includegraphics[width=0.24\linewidth]{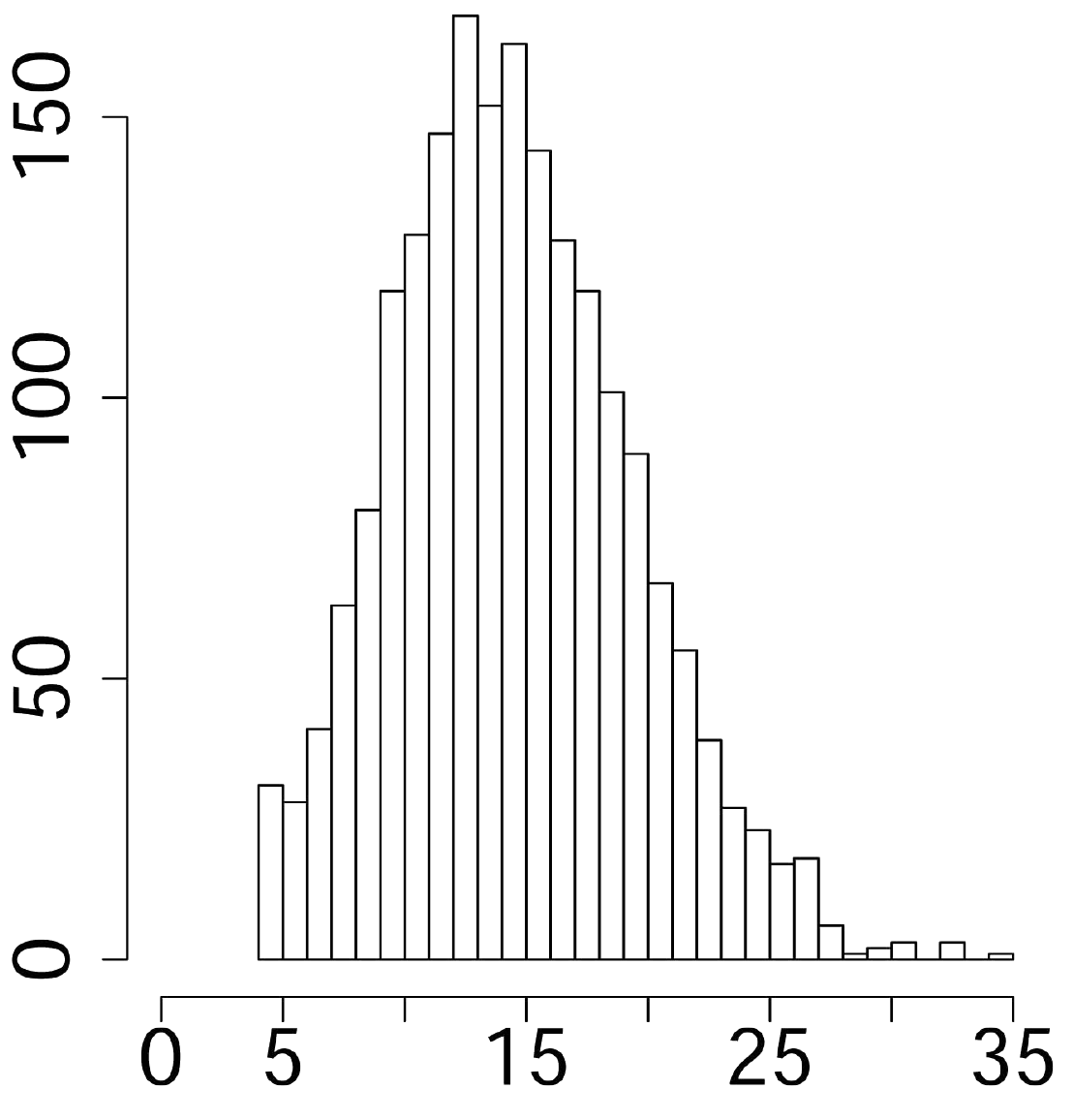}
\includegraphics[width=0.24\linewidth]{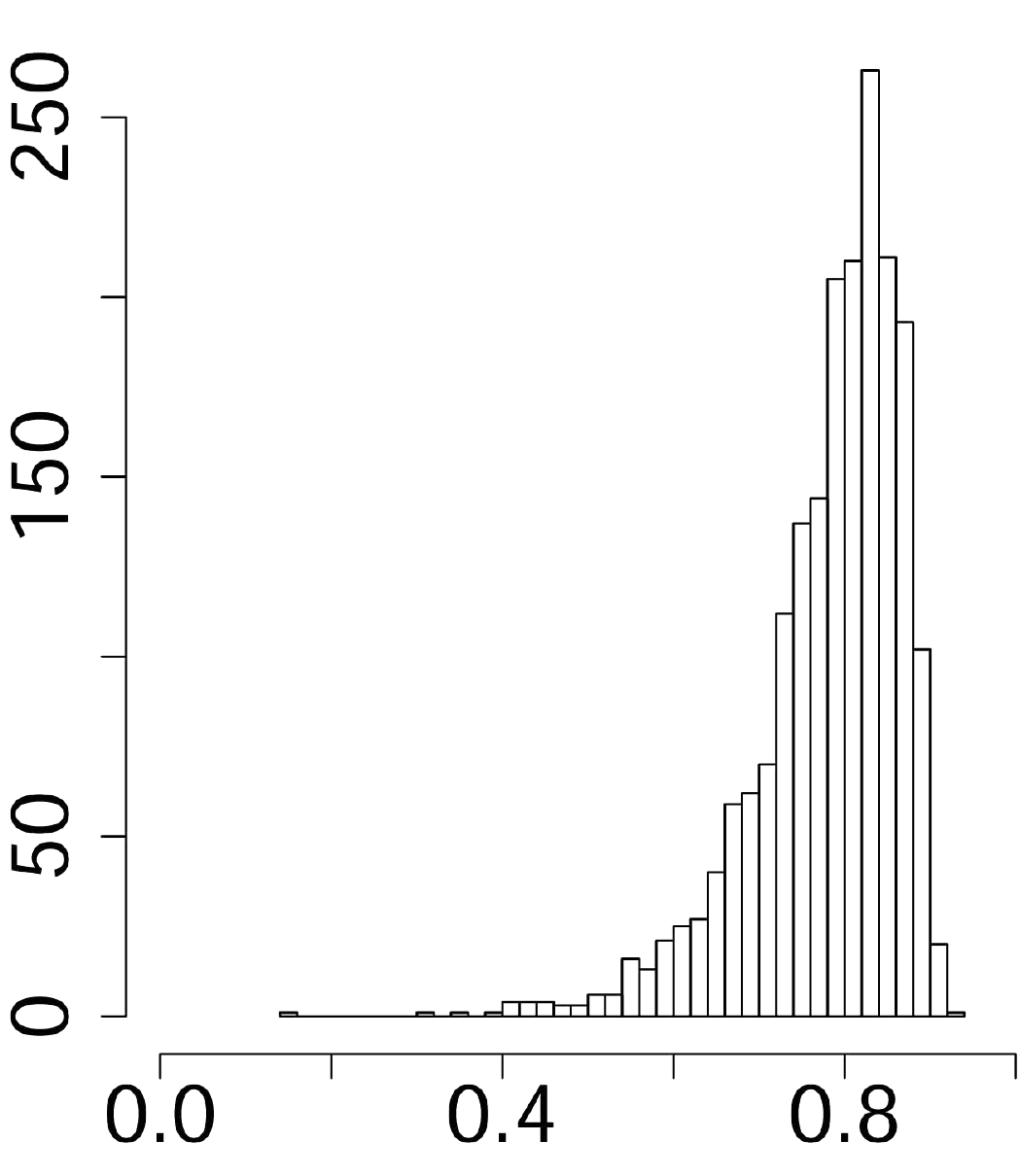}
\includegraphics[width=0.24\linewidth]{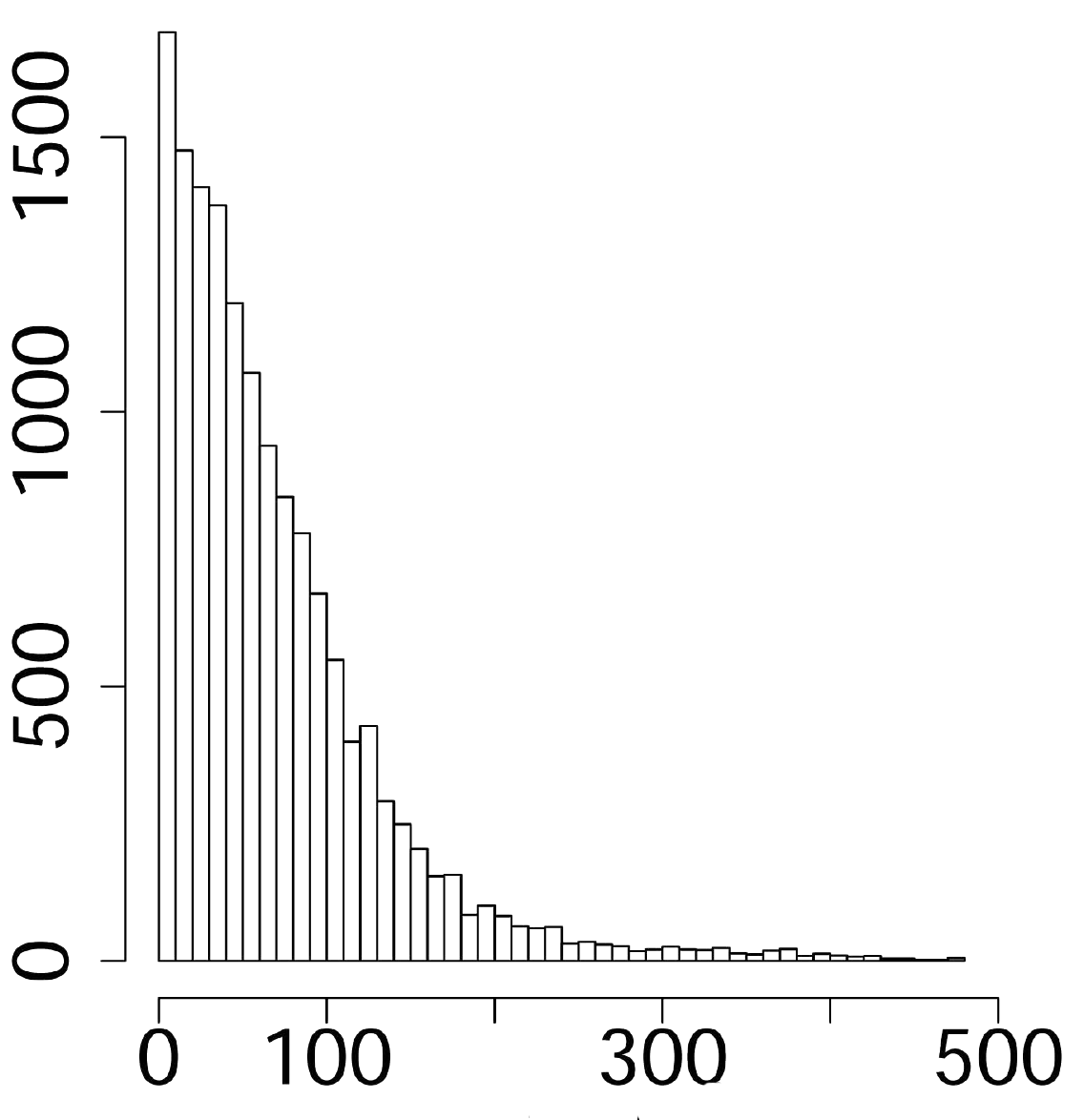}
\caption{From left to right, histograms of cell volume, number of faces, sphericity and the absolute difference in volumes for neighbouring cells. The unit for volume and difference in volume is $\mathrm \mu m^3$.} \label{fig:hist}
\end{figure}

Figure~\ref{fig:hist} depicts histograms of some cell characteristics: volume of a cell (abbreviated as `vol'); number of faces in a cell (`nof');  sphericity of a cell (`spher') as defined by
$$\text{spher} = \frac{\pi^{1/3}(6\cdot \text{vol})^{2/3}}{\text{surf}}$$
where `surf' means surface area of a cell; and
 absolute difference in volume `dvol' for two neighbouring cells (i.e., they share a face). 
Note that $0<\text{spher}\le 1$ with the value 1 corresponding to a sphere.   
 Figure~\ref{fig:hist} shows that
 many cells are small, the distribution of nof is rather symmetric and ranges from 4 (the  smallest possible value) to 35, many cells are rather spherical and many pairs of neighbouring cells have a range of differences in cell sizes in the same order as the cell size itself.
 

The upper right triangle in Table~\ref{corr} shows the empirical correlations for the cell characteristics vol, surf, nof and spher together with a new cell characteristic, namely total edge length in a cell (`tel'). The lower left triangle in the table shows the empirical correlations for dvol and three other face characteristics: area of a face (`farea'), perimeter of a face (`fper') and number of edges in a face (`fnoe'). 
We see that all correlations are high except for dvol which is nearly uncorrelated to any other face characteristic. In particular vol, surf and tel are highly correlated, and farea and fper are highly correlated (correlations $>0.9$).  


\begin{table}[h!]
\center
\scriptsize
\begin{tabular}{cccccc}
\hline\noalign{\smallskip}
 vol & 0.971 & 0.938 & 0.841 & 0.680  \\ 
 farea & surf & 0.974 & 0.874 & 0.737   \\ 
 0.923 & fper & tel & 0.941 & 0.793  \\ 
 0.751 & 0.751 & fnoe & nof & 0.754  \\ 
 0.074 & 0.062 & 0.025 & dvol & spher  \\ 
\noalign{\smallskip}\hline
\end{tabular}
\caption{\label{corr}Correlations of tessellation characteristics as defined in the text.} 
\end{table}

\begin{figure}[h!]
\center
\includegraphics[width=0.75\linewidth]{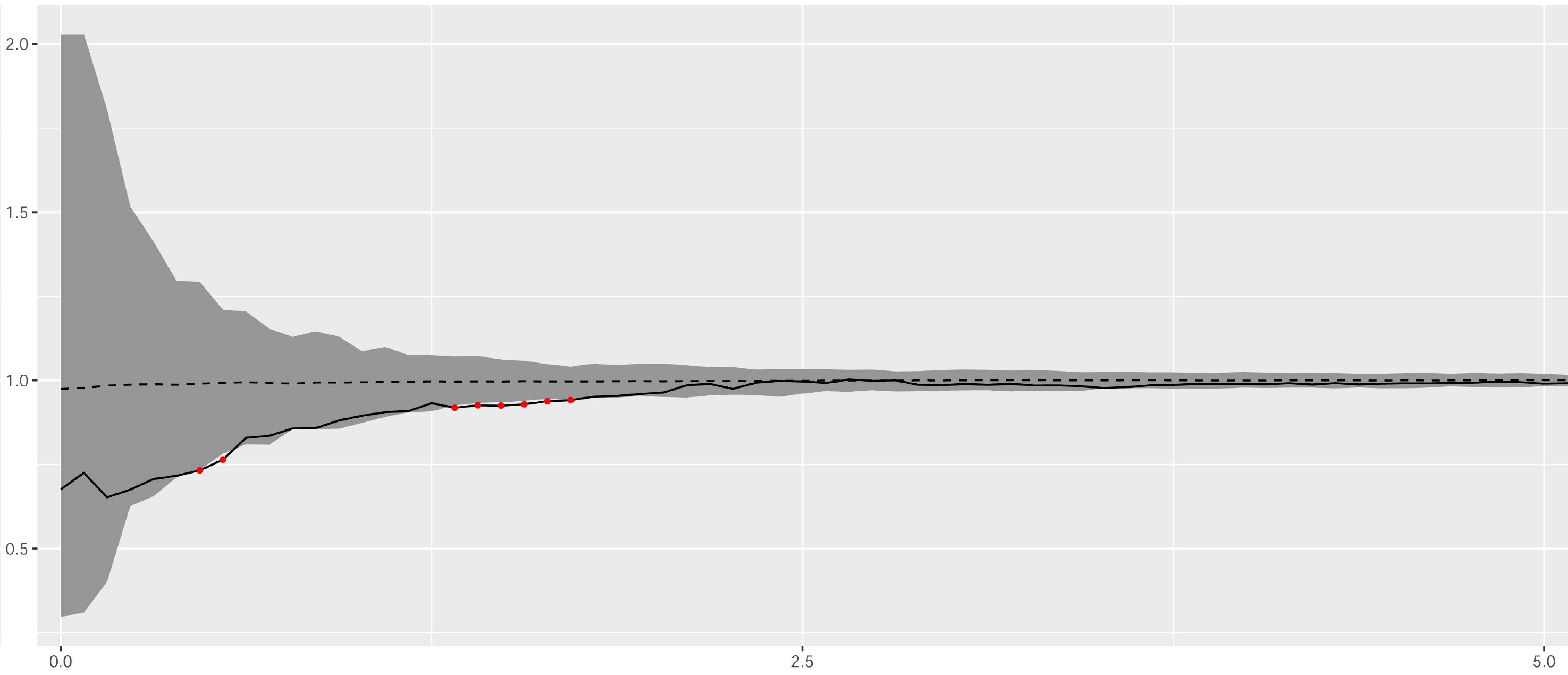}
\caption{The empirical mark correlation function (the solid line) and a 95\%-global envelope (the grey region) obtained by permuting the radii when fixing the points. The dashed line is the average of the simulated mark correlation functions 
and the dots indicate when the empirical function is outside the envelope.}
\label{fig:KMM}
\end{figure}

To see if there is dependence between $\x_n$ and $\rr_n$, we follow \cite{stoyan2021} and consider the empirical mark correlation function. We perform a permutation test based on the global area rank envelope test, 
permuting 1000 times the marks/radii when $\x_n$ is fixed and calculating the empirical mark correlation function each time.  Figure~\ref{fig:KMM} shows the function based on the data: it falls outside the 95\%-global envelope on two intervals, and the corresponding 
$p$-value obtained by the global area rank envelope test 
is 1.4\%, so we reject the null hypothesis of independence between $\x_n$ and $\rr_n$.

\section{Hierarchical model}\label{sec:hm}

This section details the two steps of our hierarchical model construction which was briefly discussed in Section~\ref{intro}. To avoid confusion, we reserve the notation $(\x_n,\rr_n)$ for the data and use the notation $(\y_m,\ttt_m)$ when we consider arguments of densities where $\y_m=\{y_1,\ldots,y_m\}\subset W$ is a finite point configuration and $\ttt_m=(t_1,\ldots,t_m)\in(0,\infty)^m$ are associated marks (if $m=0$ then $\y_0$ is the empty point configuration and $\ttt_0$ can be ignored).

\subsection{Point process models for $\x_n$}
\label{ssP}

When modelling the point pattern data set $\x_n$ as a realization of a spatial point process $\X$ on $W$,  we refer to \cite{moller2003} for background material on spatial point process models. In accordance to our observations in Section~\ref{sec:data}, we assume that the distribution of $\X$ is invariant under shifts when wrapping $W$ on a 3D torus, and a model for 
$\X$ should exhibit regularity. Pairwise interaction point processes constitute a flexible class model for regularity. Recall that $\X$ is a pairwise interaction point process on $W$ with homogeneous interaction functions (with respect to shifts on $W$ wrapped on a~3D torus) if its distribution is given by a density with respect to the unit rate Poisson process on $W$, where the density is of the form 
\[p(\y_m)=\frac{1}{Z}\beta^m\prod_{i<j}\phi(\|y_i-y_j\|_W)\]
for $m=0,1,\ldots$ and every point configuration $\y_m=\{y_1,\ldots,y_m\}\subset W$. Here 
$\beta>0$ is a parameter, $\|\cdot\|_W$ denotes shortest distance on $W$ wrapped on a~3D torus, $\phi\ge0$ is a so-called interaction function and $Z$ is a normalising constant which depends on $\beta$ and $\phi$. In order to obtain a well-defined density, a further condition on the interaction function is needed; it suffices to assume that $0\le \phi\le1$ in which case $0<Z<\infty$.


As a parametric approximation of such a pairwise interaction point process we consider a multiscale process \citep{penttinen1984}: for $d=1,2,\ldots$, let $\mathcal M_{d}$ denote the model class 
given by densities of the form
\begin{equation} \label{PIPP}
p(\y_m) \propto \beta^m \Pi_{i=1}^{d-1} \gamma_i^{\sum_{j<k} \mathbb{I}_{[\delta_{i-1} < \|y_j-y_k\|_W \leq \delta_i]}}
\end{equation}
where $\beta>0$, $0\le\gamma_1\le1$, $\ldots$, $0\le\gamma_{d-1}\le1$ and $0<\delta_1<\ldots<\delta_{d-1}$ are unknown parameters. Here we have omitted the normalising constant (which depends on all parameters), $\mathbb I_{[\cdot]}$ denotes the indicator function and we set $\delta_0=0$ and $0^0=1$. If $d=1$ we interpret the right hand side in \eqref{PIPP} as $\beta^m$. Note that $\mathcal M_1$ is just a~homogeneous Poisson process on $W$ with intensity $\beta$ and 
$\mathcal M_2$ is just a Strauss process.

Simulation under the Poisson model $\mathcal M_1$ is well-known (see e.g.\ \cite{moller2003}) 
and for simulation of the models $\mathcal M_d$, $d\ge2$, we used the birth-death-move Metropolis-Hastings algorithm \citep{geyer1994,moller2003}. 

\subsection{Exponential tessellation models for $\rr_n$ given $\x_n$}\label{s:m2}

When modelling the observed radii $\rr_n$ as a realisation of an $n$-dimensional vector, we condition on $\X=\x_n$ and consider a conditional probability  density function (pdf) $p(r_1,\ldots,r_n\mid x_1,\ldots,x_n)$ on $(0,\infty)^n$. 
More precisely we assume $\RR$ is a random vector (of random length) which conditioned on $\X=\y_m$ has a~conditional pdf $p(\ttt_m\mid \y_m)$ which is zero whenever $C(y_j,t_j\mid \y_m^*,\ttt_m^*)=\emptyset$ for some $j\in\{1,\ldots,m\}$.
Furthermore, in order to work with a well-defined conditional pdf in \eqref{MR} below we assume a mark space $\M = [0,6]$ so that $p(\ttt_m\mid \y_m)=0$ if $\ttt_m\not\in \M^m$; this is in accordance to our data, where $\min\rr_n\approx 1.00$ and $\max\rr_n\approx5.15$.

We now give a general exponential family form of the conditional pdf where $q\in\{1,2,\ldots\}$ is the dimension, $\theta=(\theta_1,\ldots,\theta_q)$ denotes the canonical parameter and $H=(H_1,\ldots,H_q)$ the canonical sufficient statistic: for $\ttt_m\in\M^m$,
\begin{equation}  
p(\ttt_m\mid \y_m) \propto
\mathbb{I}_{[C(y_j,t_j\mid \y_m^*,\ttt_m^*)\not=\emptyset,\, j=1,\ldots,m]} \exp\left(\sum_{i=1}^q\theta_i H_i(\y_m,\ttt_m)\right). \label{MR}
\end{equation}
The idea is to let each $H_i(\y_m,\ttt_m)$ depend on either the radii, or
tessellation characteristics of the cells $C(y_j,t_j\mid\y_m^*,\ttt_m^*)$ with $j=1,\ldots,m$ or interactions between these cells (here $\y_m^*$ is defined in a similar way as $\x^*$, i.e., in the right hand side of \eqref{e:jm0} $(\x,\rr)$ is replaced by $(\y_m,\ttt_m)$). 
Specifically, in  Section~\ref{sec:modsel} we consider 
the following cases (a)--(e), using similar abbreviations as in Section~\ref{sec:data}, for short writing $H_i$ for $H_i(\y_m,\ttt_m)$ and considering in (b)--(d) a sum over the cells $C(y_j,t_j\mid\y_m^*,\ttt_m^*)$, $j=1,\ldots,m$, and in (e) a sum over all unordered pairs of cells sharing a face: 
\begin{enumerate}
\item[(a)] including both $H_i = \sum_{j=1}^m \log{\frac{t_j}{6}}$ and $H_{i^{\prime}} = \sum_{j=1}^m \log{\left(1-\frac{t_j}{6}\right)}$ (with $i\neq i^{\prime}$) somehow corresponds to a scaled beta distribution for the radii if no other terms are included in \eqref{MR} -- `somehow' because it is not exactly a beta distribution since $p(\ttt_m\mid \y_m)=0$ if $C(y_j,t_j\mid \y_m^*,\ttt_m^*)=\emptyset$ for some $j\in\{1,\ldots,m\}$; 
\item[(b)] $H_i=\sum{\mathrm{nof}}$ 
is twice the total number of faces;
\item[(c)] $H_i=\sum{\mathrm{surf}}$ is twice the total surface area of faces;
\item[(d)] $H_i=\sum{\mathrm{vol}}^2$ is the sum of squared volumes of cells;
\item[(e)] $H_i=\sum{\mathrm{dvol}}$ is the sum of difference in volumes of two cells which share a face.
\end{enumerate}
 The density in \eqref{MR} is then well-defined for all $\theta\in\mathbb R^q$ provided in case of (a) we have $\theta_i>-1$ and $\theta_{i^{\prime}}>-1$; this 
follows since the mark space is bounded. Moreover,  
for simulation under \eqref{MR} we use a Metropolis within Gibbs algorithm where we alternate between updating from the 
conditional densities
\begin{equation}\label{e:jm1}
p(t_j\mid t_k,\,k\not=j,\, y_1,\ldots,y_m)\propto p(t_1,\ldots,t_m\mid y_1,\ldots,y_m),\quad j=1,\ldots,m,
\end{equation}
using a Metropolis algorithm with a normal proposal.

\subsection{Estimation}\label{sec:est}

We assume that the unknown parameter $(\theta_1,\ldots,\theta_q)$ in \eqref{MR} varies independently of the parameters in \eqref{PIPP}. 
Then parameter estimation can be done in two steps and the results will not depend on each other. 

To avoid calculating the intractable normalising constants appearing in \eqref{PIPP} and \eqref{MR}, we use
 maximum pseudolikelihood methods \citep{besag1974,besag1977a,besag1982} 
  in both cases  
  where a few remarks are in order: For the definition of pseudolikelihood function in connection to \eqref{PIPP}, we refer to \cite{jensen1991}; see also \ref{ap:4}.
  For fixed $(\delta_1,\ldots,\delta_d)$, \eqref{PIPP} is an exponential family model with canonical parameter $\vartheta=(\log\beta,\log\gamma_1,\ldots,\log\gamma_d)$ and so the log pseudolikelihood function is a concave function of $\vartheta$ \citep{jensen1991} and  
   we can find a partial maximum pseudolikelihood estimate (MPLE) of $\vartheta$ using the Newton-Raphson algorithm. Hence we obtain a profile log pseudolikelihood which is maximised with respect to $(\delta_1,\ldots,\delta_d)$ defined over a $d$-dimensional grid, thereby providing the final MPLE. Furthermore, when considering \eqref{MR} the pseudolikelihood for $\theta$ is defined by the product of the conditional densities of each $r_j$ given the $r_k$ with $k\not=j$, cf.\ \eqref{e:jm1}. Since \eqref{MR} is also an exponential family, the log pseudolikelihood is concave and then again Newton-Raphson can be used for finding the MPLE. Finally, we calculate various integrals which appear in the pseudolikelihoods (in case of \eqref{PIPP}, see \cite{jensen1991}; in case of \eqref{MR}, we consider the integral of the right hand side of \eqref{e:jm1} with respect to $t_j$) by numerical methods.

\section{Model selection and final joint model}\label{sec:modsel}

This section details the model selection procedure briefly described in Section~\ref{s:intro.2}
and which
leads to 
our final joint model for $\x_n$ and $\rr_n$. The classical goodness of fit procedures are based on measures given by maximizing likelihood functions and accounting for model complexity by including a penalty in terms of the number of parameters (e.g.\ AIC and BIC criteria). In \cite{sed2018} the Vapnik-Chervonenkis theory is used instead and a criterion based on residual sum of squares is developed for partially ordered sets of deterministic tessellation models. Since we deal with stochastic tessellation models but our likelihood functions are intractable to maximize, we present another approach that corresponds to modern trends in spatial statistics. It is based on how well geometrical tessellation characteristics are described when comparing fitted models, where we account for mutual correlations among the characteristics and we consider the maximum of pseudolikelihood functions when comparing models with the same number of parameters.  

\subsection{The fitted model for $\x_n$}\label{s:fit1}

\begin{figure}[h!]
\center
\includegraphics[width=0.3\linewidth]{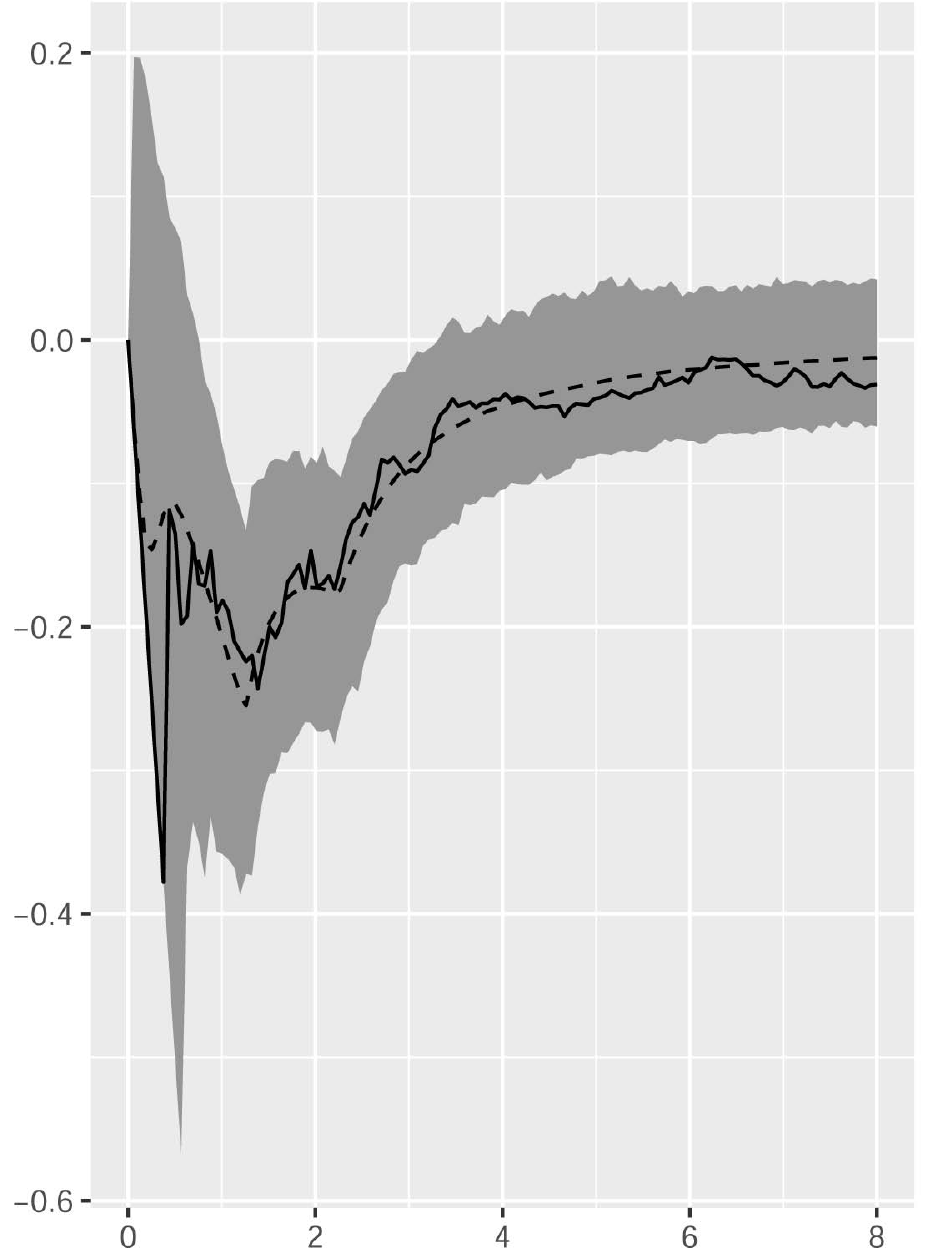}
\includegraphics[width=0.3\linewidth]{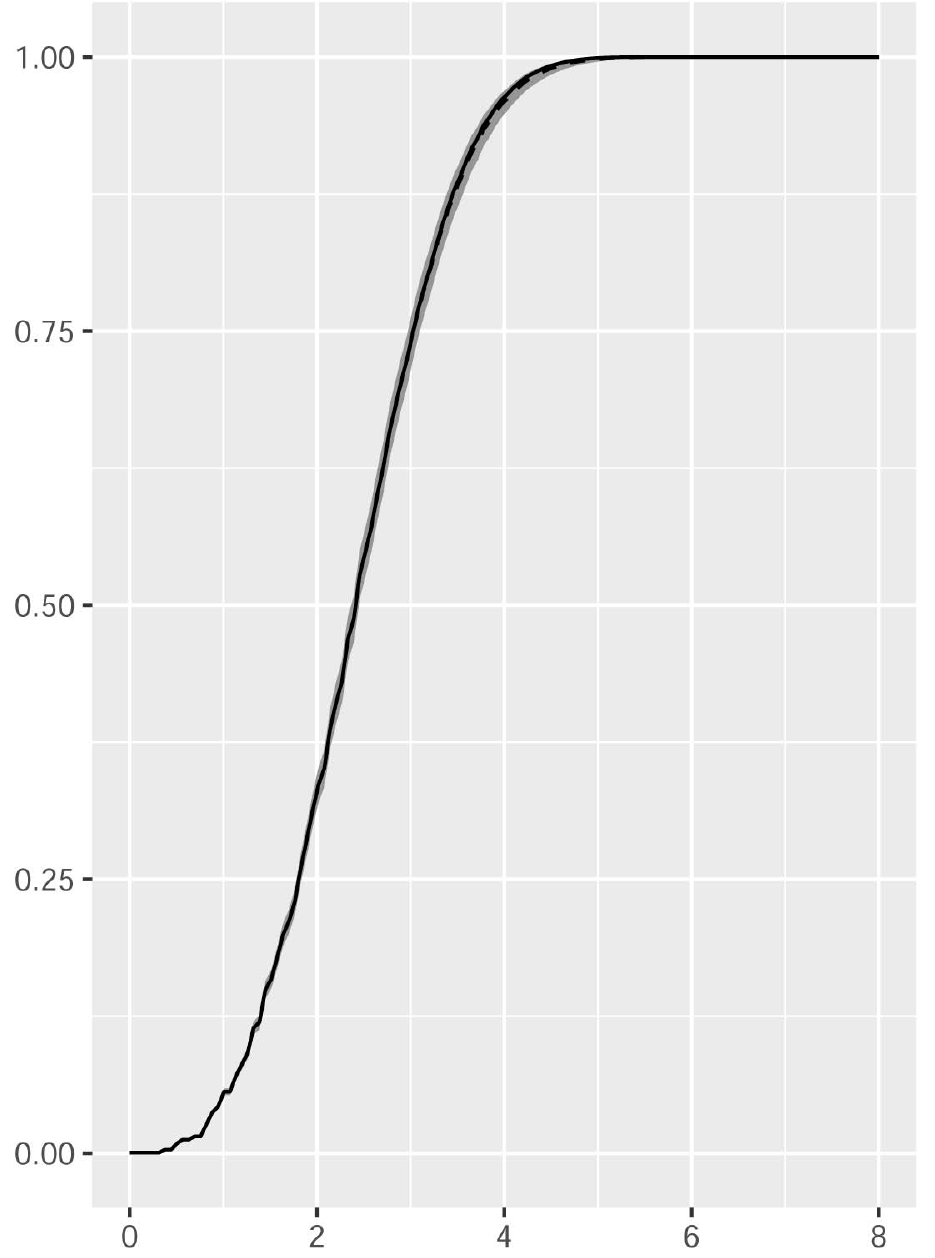}
\includegraphics[width=0.3\linewidth]{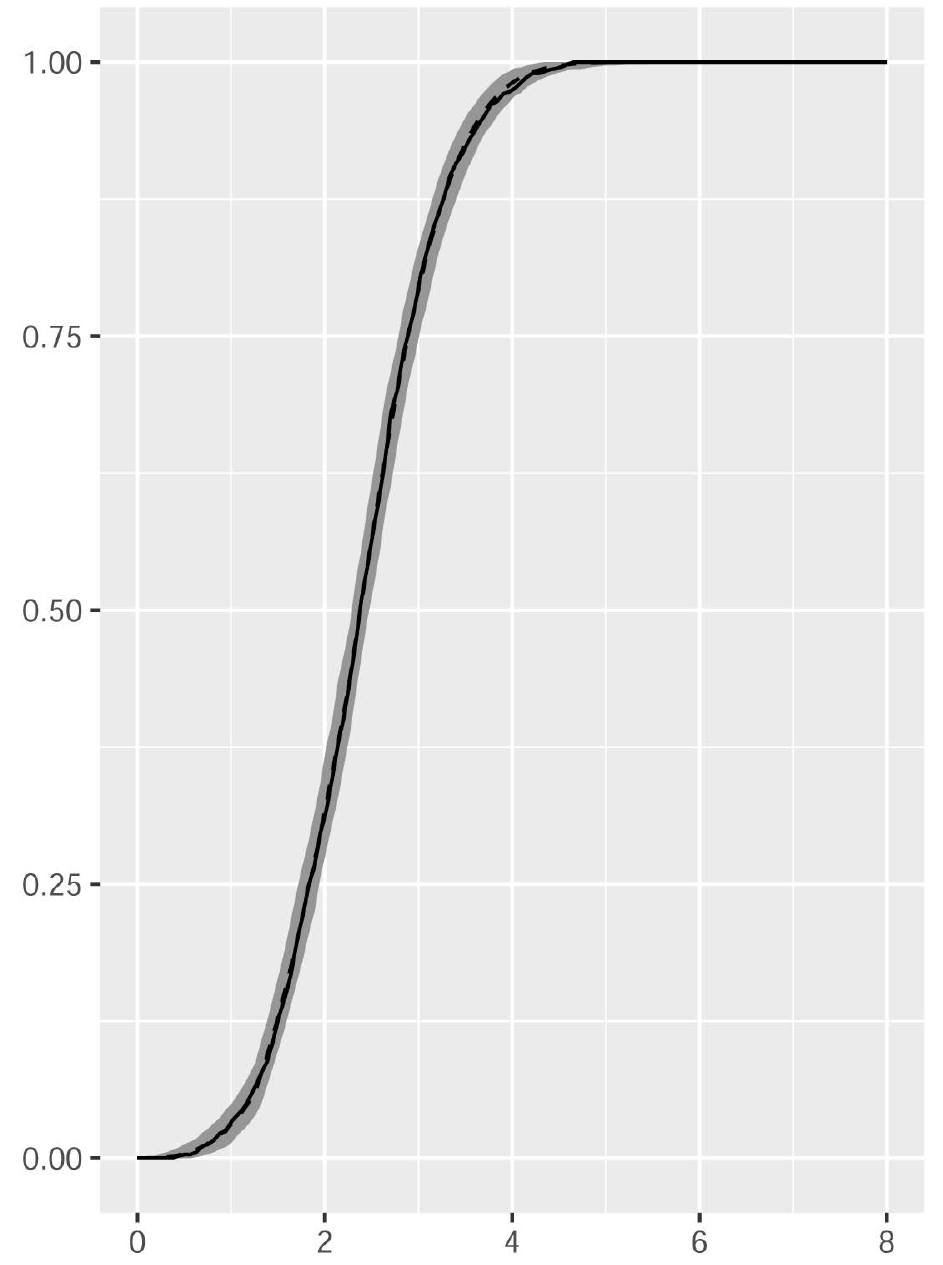}
\caption{From left to right, empirical functional summary statistics $\hat L(t)-t$, $\hat F(t)$ and $\hat G(t)$ (solid lines) and simulated 95\%-global envelope (grey regions) obtained under a fitted multiscale process with $d=3$. Dashed lines are averages of the simulated functional summary statistics.}
\label{fig:PP2_GET}
\end{figure}

For the point pattern $\x_n$ and $d=1,2,\ldots$, we select the first model $\mathcal M_d$ which provides a satisfactory fit when considering global envelopes and global area rank envelope test in the same way as in Section~\ref{sec:data}.
As noticed in Section~\ref{sec:data}, the Poisson model $\mathcal M_1$ is not providing a satisfactory fit and as discussed in \ref{ap:3}
this is also the case for the Strauss model $\mathcal M_2$. The
 first model which is not significant at level 5\% is $\mathcal M_3$: Figure~\ref{fig:PP2_GET} shows that the empirical functional summary statistics are within the 95\%-global envelope produced in a similar way as in Figure~\ref{fig:PP0_GET} but using simulations under the fitted multiscale process with $d=3$. The corresponding 
$p$-value obtained by the global area rank envelope test 
 is $18.8\%$, and the maximum pseudolikelihood estimates (see Section~\ref{sec:est}) are $\hat{\beta} = 0.0168$, $\hat{\gamma}_1 = 0.5328$, $\hat{\gamma}_2 = 0.8432$, $\hat{\delta}_1 = 1.25$ and $\hat{\delta}_2 = 2.25$.


\subsection{The fitted model for $\rr_n$ conditioned on $\x_n$}

Now, having fitted the model $\mathcal M_3$ for $\mathbf x_n$ it remains to obtain a model for $\rr_n$ conditioned on $\x_n$ where we use a model selection procedure as follows. We consider a list of tessellation characteristics
\begin{equation} \label{list}
{\cal L} = \{ \text{nof}, \text{vol}, \text{surf}, \text{tel}, \text{spher}, \text{dvol} \}
\end{equation}
when creating and evaluating more and more complex models as given by \eqref{MR} and (a)--(e) in Section~\ref{s:m2}. When comparing fitted models of the same dimension $q$, we select the one with the highest value of the maximized log pseudolikelihood function. The selected model is then evaluated by a global area rank envelope test based on kernel smoothed densities of the empirical distributions for the six tessellation characteristics in $\mathcal L$, using $499$ simulations of the joint model for $\x_n$ and $\rr_n$. Here  
we concatenate the six densities and use the \texttt{R}-package \texttt{GET} \citep{myllymaki2019}, and we simulate from the joint model rather than from the conditional model of $\mathbf r_n$ given $\mathbf x_n$, since our aim is to replace expensive laboratory experiments with simulations from the joint model. 
Therefore, in Table~\ref{modsel}, we also evaluate our fitted models by comparing empirical (column `data') and simulated means and standard deviations of the tessellation characteristics in $\mathcal L$, using 100 simulations of first $\y_m$ and second under the different fitted models of $\ttt_m$ conditioned on $\y_m$. The table also shows the (signed) difference between the empirical and the simulated values divided by the empirical values (the values given in percentages) -- we refer to such a value as a~deviation.

First, we consider the empirical distribution of the radii:
As the histogram in Figure~\ref{fig:radii} looks like a beta distribution, we first propose in \eqref{MR} only to include the terms in (a) so that $q=2$, $H_1=\sum_{j=1}^m\log\tfrac{t_j}{6}$ and $H_2=\sum_{j=1}^m\log\left(1-\tfrac{t_j}{6}\right)$ -- we refer to this model as `beta'.
The column `beta' in Table~\ref{modsel} shows 
that except for dvol the means match the data well, and the deviations of standard deviations vary by $12$ to $24$ percent. However, the fitted `beta' model was highly significant when evaluated by the global area rank envelope test.  

\begin{table}[h!]
\center
\scriptsize
\begin{tabular}{ccccc|c}
\hline\noalign{\smallskip}
&   & models & & & data\\
&    &  beta & ${\mathrm{beta}} + {\mathrm{dvol}}$ & ${\mathrm{beta}} + {\mathrm{nof}} + {\mathrm{dvol}}$ & \\
\noalign{\smallskip}\hline\noalign{\smallskip}
nof & mean  &  14.82 & 14.87 & 14.95 & 14.98 \\
&   & -1\% & -1\% & $<$-1\% & \\
& sd  & 5.58 & 5.38 & 5.20 & 4.92 \\
&   & +13\% & +9\%  & +6\% &  \\
vol & mean & 70.65 & 70.65 & 70.65 & 69.21 \\
&   & +2\% & +2\% & +2\% &  \\
& sd  &  65.65 & 63.42 & 62.01 & 58.89 \\
&   & +12\% &  +8\% & +5\% &  \\
surf & mean & 91.14 & 91.53 & 93.58 & 92.46 \\
&   & -1\% &  -1\% & +1\% & \\
& sd  & 57.26 & 54.07 & 51.50 & 47.86 \\
&   & +20\% & +13\% & +8\% &  \\
tel & mean & 66.16 & 66.12 & 67.69 & 67.92 \\
&   & -3\% &  -3\% & $<$-1\% &   \\
& sd  & 31.99 & 30.86 & 29.52 & 27.38 \\
&   & +17\% & +13\% & +8\% &  \\
spher & mean  & 0.76 & 0.76 & 0.77 & 0.78 \\
&   & -3\% & -3\% & -1\% &  \\
& sd  & 0.11 & 0.10 & 0.10 & 0.087 \\
&   & +24\% & +19\% & +16\% &  \\
dvol & mean  & 79.87 & 73.72 & 69.82 & 68.89 \\
&   & +16\% &  +7\% & +1\% & \\
& sd  & 72.75 & 68.57 & 62.78 & 65.04 \\
&   & +12\% & +5\% & -4\% &  \\
\noalign{\smallskip}\hline
\end{tabular}
\caption{\label{modsel}Means and standard deviations of the tessellation characteristics given in \eqref{list} and as obtained by simulations under various joint models 
and from the data. {Deviations are given in percentages.} See the text for details. 
} 
\end{table} 

Second, we expand the model by including one of the terms $H_3=\sum{\mathrm{vol}}^2$, $\sum{\mathrm{nof}}$, $\sum{\mathrm{surf}}$ or $\sum{\mathrm{dvol}}$ so that $q=3$. Here, we do not include $\sum{\mathrm{vol}}$, since this is a constant; or $\sum{\mathrm{tel}}$, since the correlation coefficient between vol and tel for each cell is close to 1; or $\sum{\mathrm{spher}}$, since by definition spher is given by vol and surf for each cell. Table~\ref{tab:score} shows that the model with $H_3=\sum{\mathrm{dvol}}$ has the largest maximized log pseudolikelihood function. For this model, comparing the columns `beta' and `beta+dvol' in Table~\ref{modsel} we obtain now  better results for dvol. Moreover, all standard deviations are now reduced, and the $p$-value based on global area rank envelope test is 9.8\%. 

\begin{figure}[h!]
\includegraphics[width=1\linewidth]{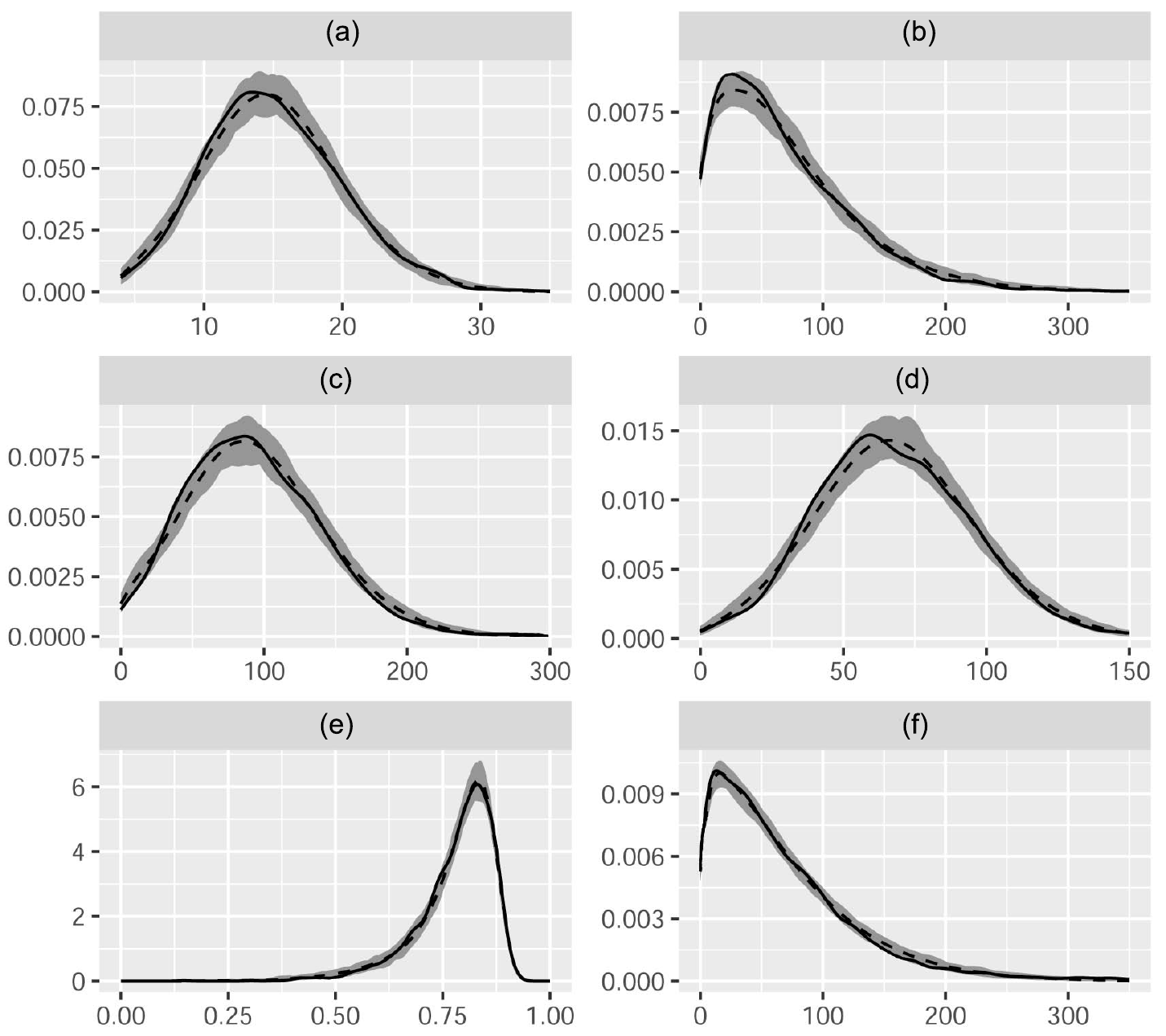}
\caption{
Estimated densities of tessellation characteristics, namely nof (a), vol (b), surf (c), tel (d), spher (e) and dvol (f). The solid lines are the functions based on the data $(\x_n,\rr_n)$ and the grey regions are 95\%-global envelopes under the fitted model `beta+nof+dvol'. Dashed lines are averages of the simulated densities.
} 
\label{fig:rad2_GET} 
\end{figure}

Third, we investigate the effect of expanding the model `beta' with any two of the four terms $\sum{\mathrm{vol}}^2$, $\sum{\mathrm{nof}}$, $\sum{\mathrm{surf}}$ and $\sum{\mathrm{dvol}}$ so that $q=4$. Table~\ref{tab:score} shows that `beta+nof+dvol' provides the best fit according to the maximized log pseudolikelihood function. For this model, the $p$-value based on global area rank envelope test is 10.6\% which is slightly larger than the $p$-value of 9.8\% for the model `beta+dvol'. Table~\ref{modsel} 
shows an improvement for both the mean values and the standard deviations when comparing `beta+nof+dvol' with `beta+dvol'. Figure~\ref{fig:rad2_GET} shows empirical 
kernel estimates of the densities for the six characteristics in $\mathcal L$ together with a  
$95\%$-global envelope obtained when concatenating all six empirical densities. 
The empirical functions 
are completely covered by the envelope.  Moreover, the maximum pseudolikehood estimates are $\hat{\theta}_1 = 4.709$, $\hat{\theta}_2 = 5.982$, $\hat{\theta}_3 = -2.376\times 10^{-1}$ and $\hat{\theta}_4 = 3.021\times 10^{-2}$. Thus under this fitted model realisations become more likely as the total number of faces decreases or the sum of differences in volumes between neighbouring cells increases (when all other terms $H_i$ in \eqref{MR} are fixed).


In a similar way, we also tried fitting models without including the terms in (a), thereby obtaining `nof+surf+dvol' as the final model for $q=3$ and where the $p$-value is 7.7\%. A plot similar to Figure~\ref{fig:rad2_GET} but for the fitted `nof+surf+dvol' model is given in \ref{ap:5}.

\begin{table}[h!]
\center
\scriptsize
\begin{tabular}{cc}
\hline\noalign{\smallskip}
model & $\log${\cal{PL}}$(\hat{\theta})$ \\
\noalign{\smallskip}\hline\hline\noalign{\smallskip}
${\mathrm{beta}}$ & -2532.45 \\
\noalign{\smallskip}\hline\hline\noalign{\smallskip}
${\mathrm{beta}} + {\mathrm{vol}}^2$ & -2866.39 \\
${\mathrm{beta}} + {\mathrm{nof}}$ & -2594.72 \\ 
${\mathrm{beta}} + {\mathrm{surf}}$ & -2483.98 \\
${\mathrm{beta}} + {\mathrm{dvol}}$ & -2468.76 \\
\noalign{\smallskip}\hline\hline\noalign{\smallskip}
${\mathrm{beta}} + {\mathrm{vol}}^2 + {\mathrm{nof}}$ & -2513.89 \\
${\mathrm{beta}} + {\mathrm{vol}}^2 + {\mathrm{surf}}$ & -2823.71 \\
${\mathrm{beta}} + {\mathrm{vol}}^2 + {\mathrm{dvol}}$ & -2714.10 \\
${\mathrm{beta}} + {\mathrm{nof}} + {\mathrm{surf}}$ & -2498.58 \\
${\mathrm{beta}} + {\mathrm{nof}} + {\mathrm{dvol}}$ & -2465.12 \\
${\mathrm{beta}} + {\mathrm{surf}} + {\mathrm{dvol}}$ & -2477.63 \\
\noalign{\smallskip}\hline\hline\noalign{\smallskip}

\end{tabular}
\caption{\label{tab:score} Maximized log pseudolikelihood functions for radii models conditioned on the points. The value for the `beta' model is given as a reference.
}
\end{table}

\section{Concluding remarks}\label{sec:con}

%
%
%
%
%

Our hierarchical model construction for Laguerre tessellation data sets, using a Gibbs model for the points and a conditional model for the radii given the points, may be useful in several other contexts, including 2D cases. 
As compared to the use of Gibbs models in \cite{dereudre2011} and  \cite{seitl2020}, our hierarchical approach is less time demanding: To simulate a vast number of standard point process models is a routine task which is not too much time demanding. The more exhaustive part is to simulate the radii since Laguerre cells have to be recomputed, but still in comparison with 
\cite{seitl2020} it becomes faster in our case since we condition on the points. 


%
%

For the Gibbs point process model, we considered multiscale models which approximate the class of pairwise interaction point processes and lead to a natural model selection procedure by considering an increasing number of interaction parameters. We might consider point process models with higher order interactions such as Geyer's triplet interaction process \citep{geyer1999} or the area interaction process involving interactions of all orders \citep{baddeley1995}. However, the computational complexity increases with more complex models.   

For the conditional distribution of radii given points, we have demonstrated the usefulness of exponential family models with a canonical sufficient statistic based on tessellation 
characteristics. As there are several possibilities for selecting tessellation characteristics and there is no natural ordering (as there is in the case of the multiscale model), a model selection procedure is less straightforward. However, we demonstrated how to compare fitted models of the same dimension by considering maximized log pseudolikelihood functions and how to evaluate selected fitted models by partly comparing moment properties of tessellation characteristics under simulations from the model with empirical moments and partly by considering plots of global envelopes and by evaluating their corresponding $p$-values.


%
%
%

When calculating pseudolikelihood functions, various integrals have to be evaluated, cf.\ Section~\ref{sec:est}. 
We used simple grid-based approximation methods for both the pseudolikelihood based on the points and that based on the radii given the points. A small simulation study indicated that this may cause some bias in the MPLEs (also for MPLEs of the Gibbs models in \cite{dereudre2011} and \cite{seitl2020} some bias appeared).  \cite{BadTur} noticed that bias of MPLEs for spatial point process models usually does not reflect weaknesses of the MPLE method, but is probably due to the effects of discretization of the observation window.
For 2D spatial point process models, efficient numerical methods have been developed \citep{BadTur,baddeley2014}  and \texttt{spatstat} provides useful software, but to the best of our knowledge it remains to make progress for   in general 3D point processes and for the special case of our model. We leave this important and huge task for future research.

\section*{Acknowledgements}
The research was supported by the Czech Science Foundation, project 19-04412S, and by The Danish Council for Independent Research —
Natural Sciences, grant DFF – 7014-00074 ‘Statistics for point processes in space and beyond'. We thank the referees for useful comments and suggestions.\\

\appendix

\section{Visualization of the observed Laguerre tessellation}\label{ap:1}

\begin{figure}[h!]
\center
\includegraphics[width=0.24\linewidth]{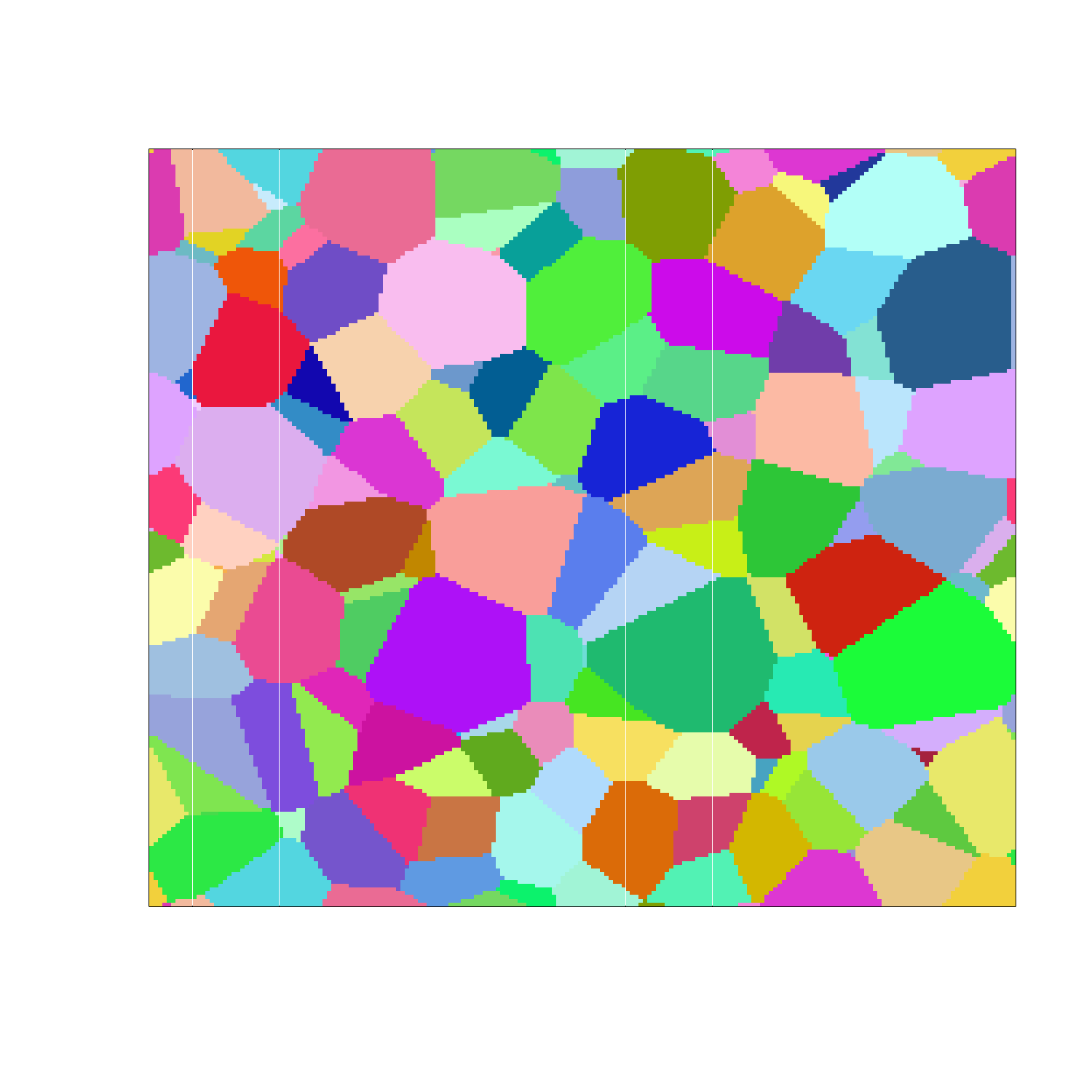}
\includegraphics[width=0.24\linewidth]{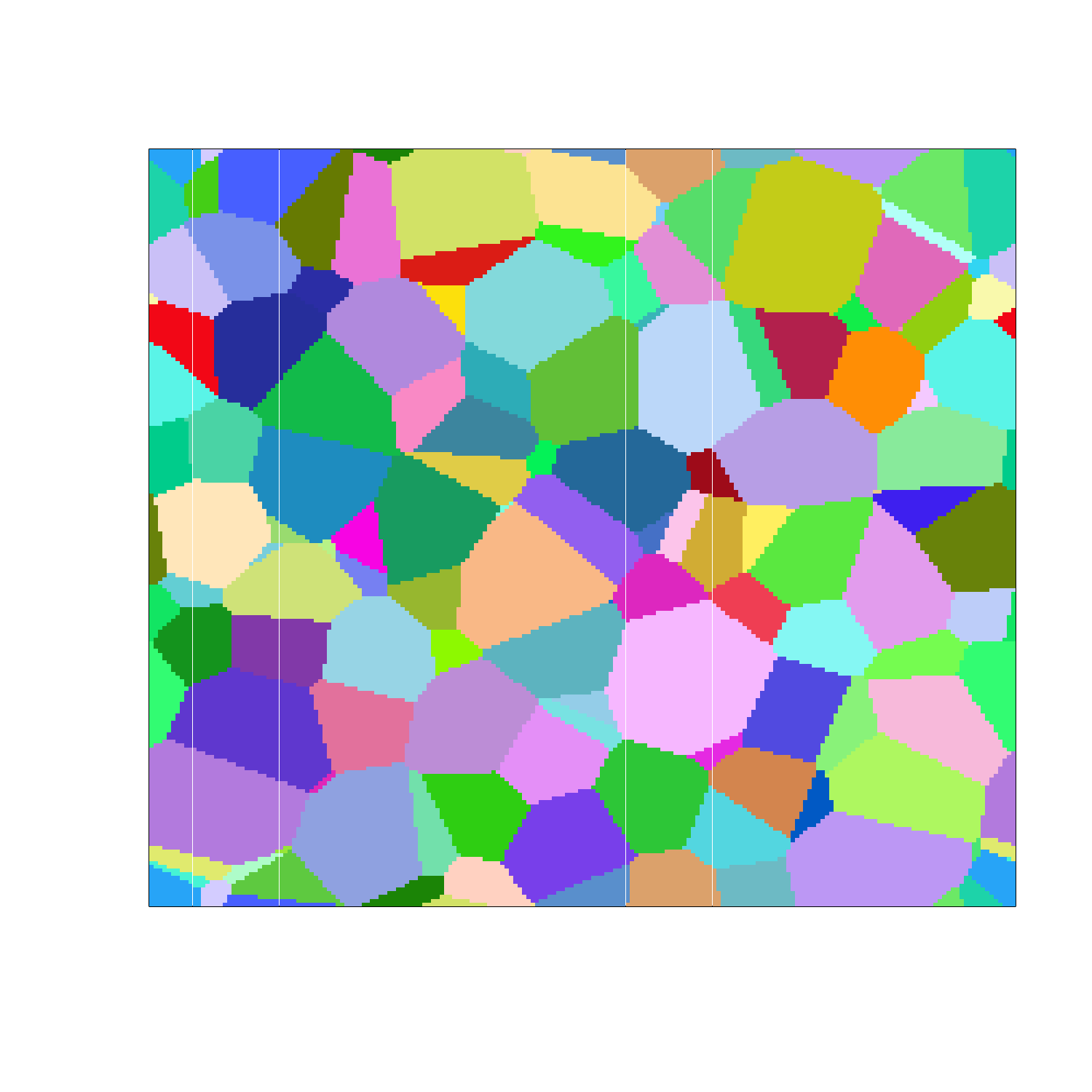}
\includegraphics[width=0.24\linewidth]{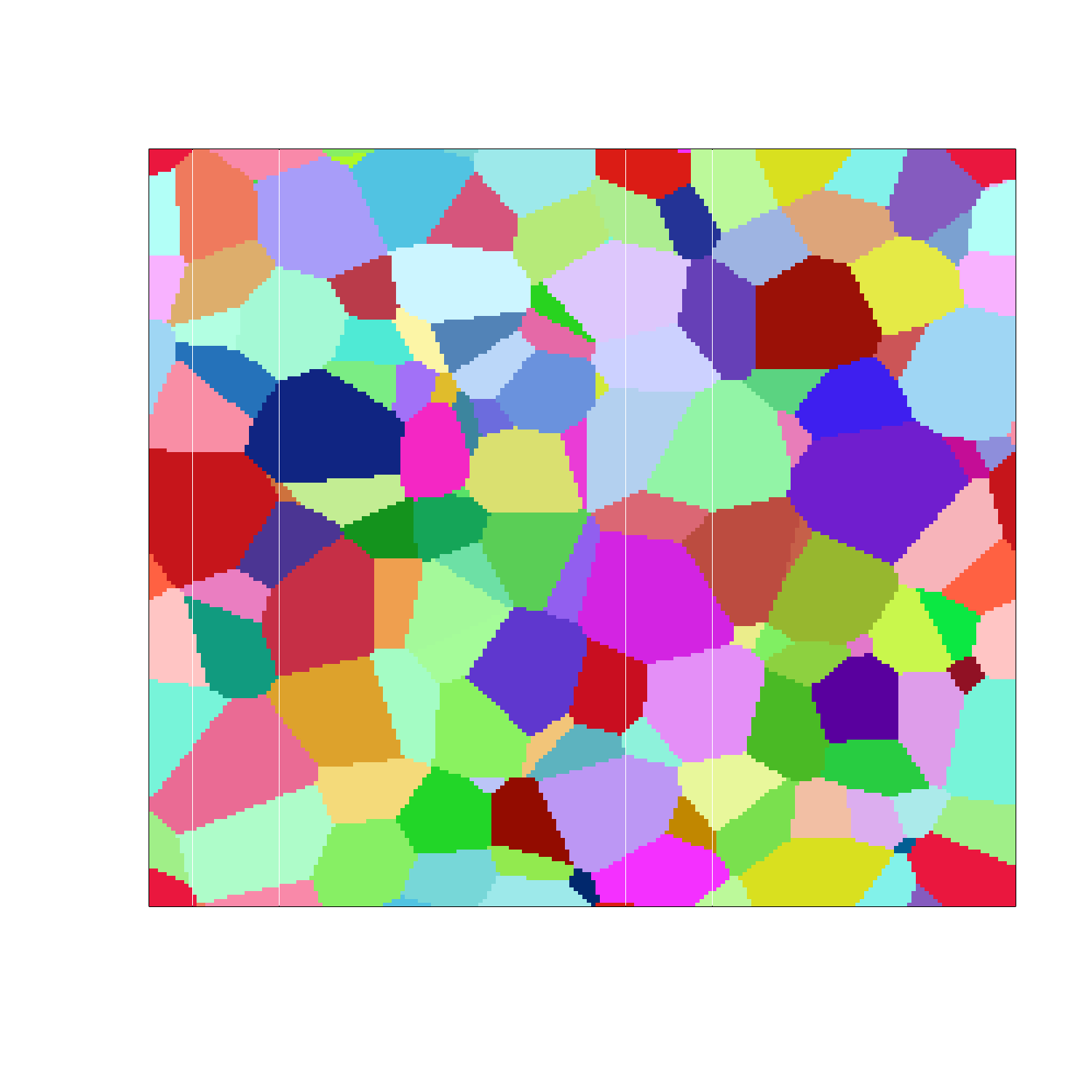}
\includegraphics[width=0.24\linewidth]{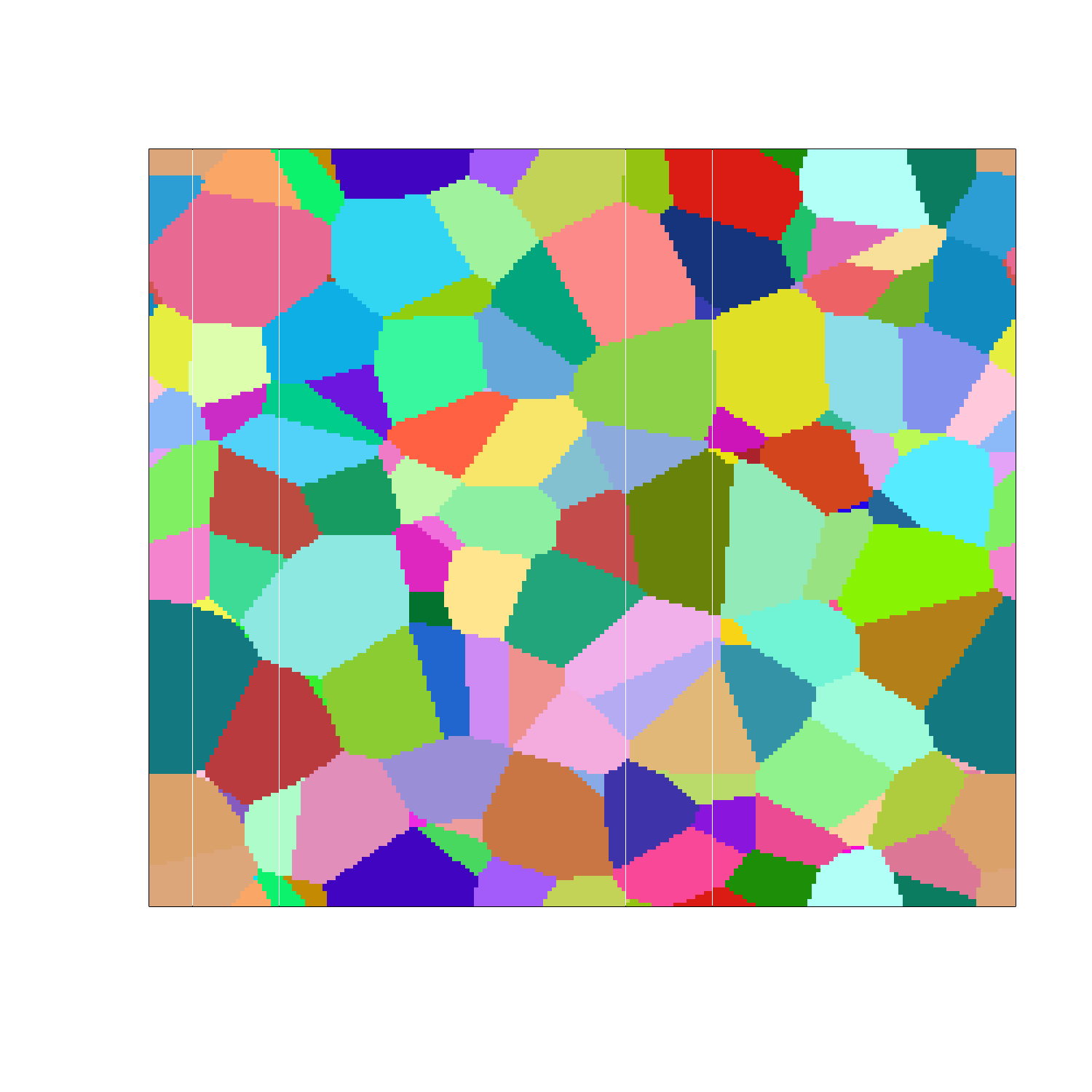}

\caption{Four slices in $xy$ plane of the observed Laguerre tessellation. } 
\label{fig:ap:image}
\end{figure}

In order to visualize the Laguerre tessellation data set described in Section~\ref{sec:data}, Figure~\ref{fig:ap:image} shows four equidistant slices perpendicular to the $z$ axis. 

\section{Homogeneity of points and radii}\label{ap:2}

\begin{figure}[h!]
\begin{minipage}{0.32\textwidth}
\includegraphics[width=1\linewidth]{Fig_H_generators_all_copy.pdf}
\end{minipage} 
\begin{minipage}{0.33\textwidth}
\includegraphics[width=1\linewidth]{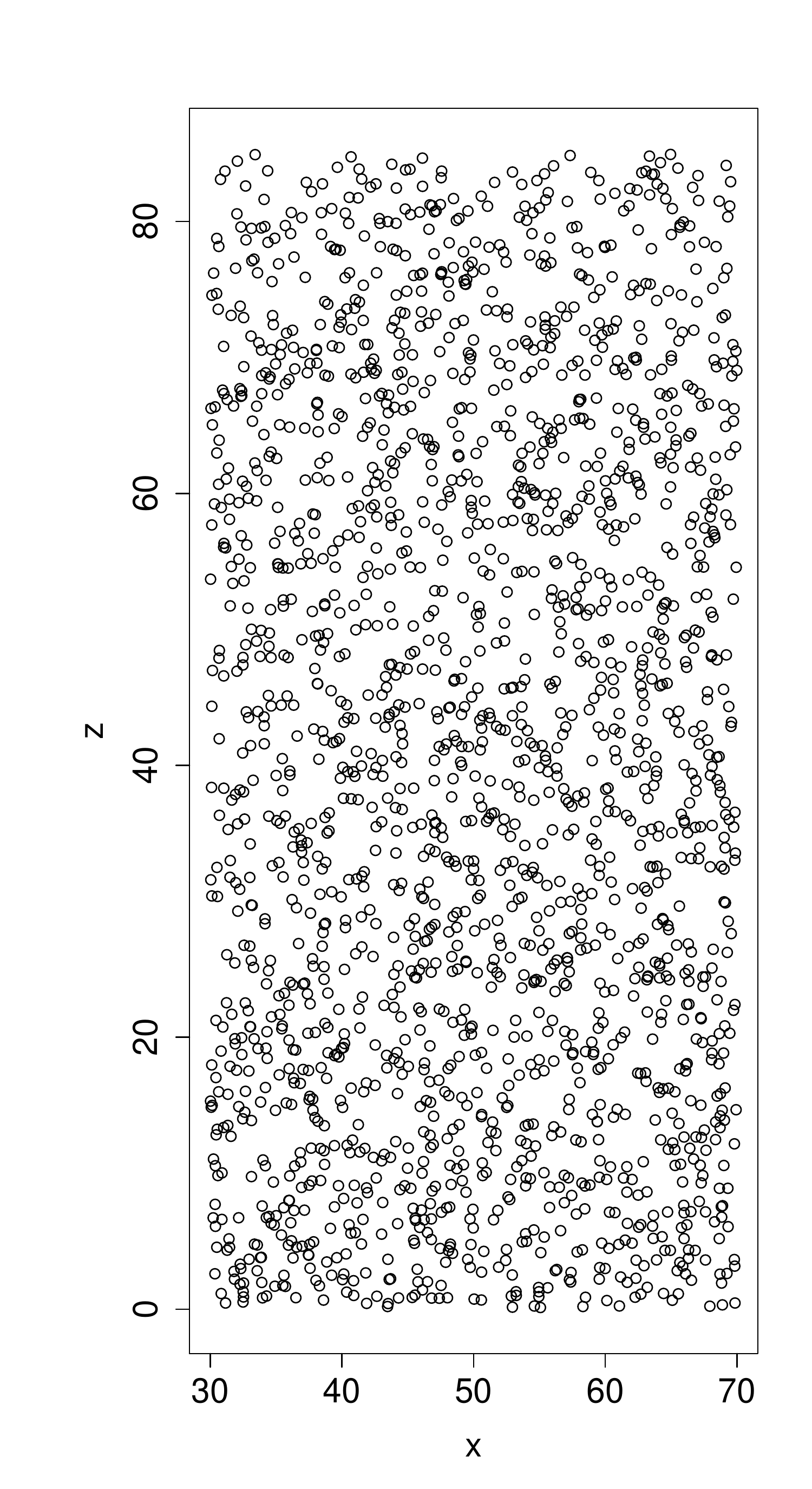}
\end{minipage} 
\begin{minipage}{0.33\textwidth}
\includegraphics[width=1\linewidth]{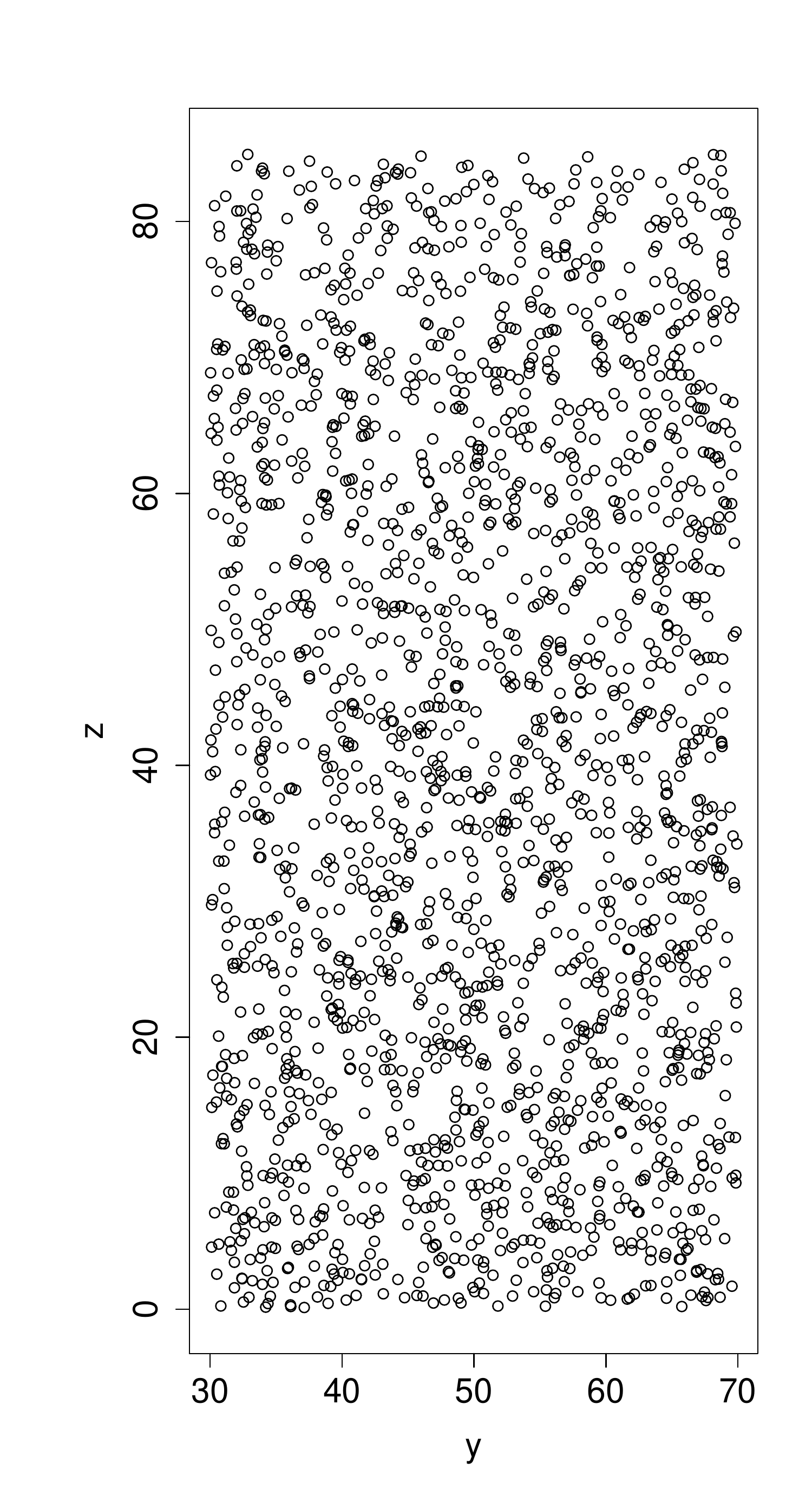}
\end{minipage} 
\begin{minipage}{0.33\textwidth}
\includegraphics[width=1\linewidth]{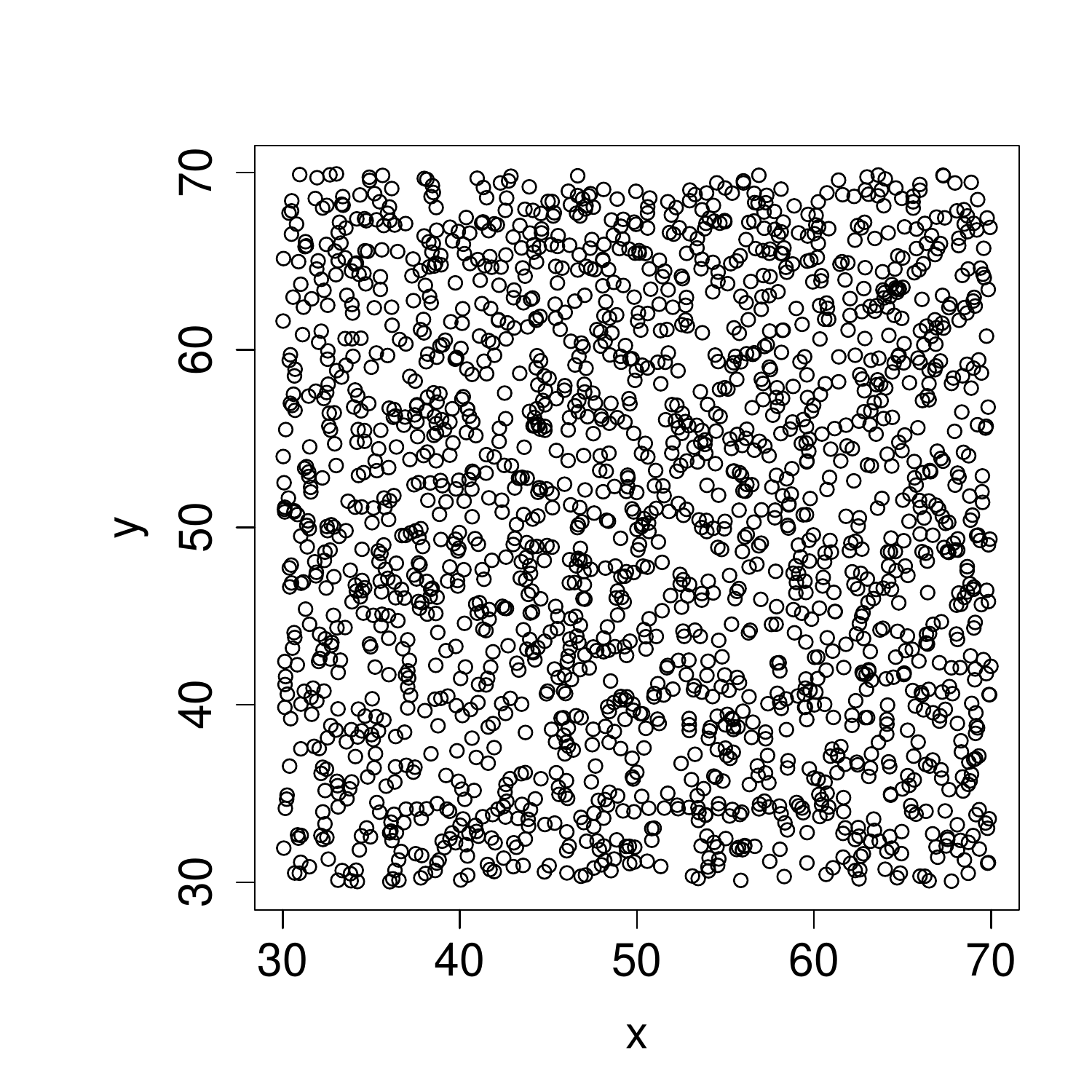}
\end{minipage} 
\begin{minipage}{0.07\textwidth}
\textcolor{white}{.}
\end{minipage} 
\begin{minipage}{0.58\textwidth}
\caption{The spatial point pattern $\x_n$ of the generators of the Laguerre tessellation after applying a periodical boundary condition together with its projections onto the $xy$, $xz$ and $yz$ planes.}
\label{fig:ap:generators}
\end{minipage} 
\end{figure}

\begin{figure}[h!]
\center
\includegraphics[width=0.4\linewidth]{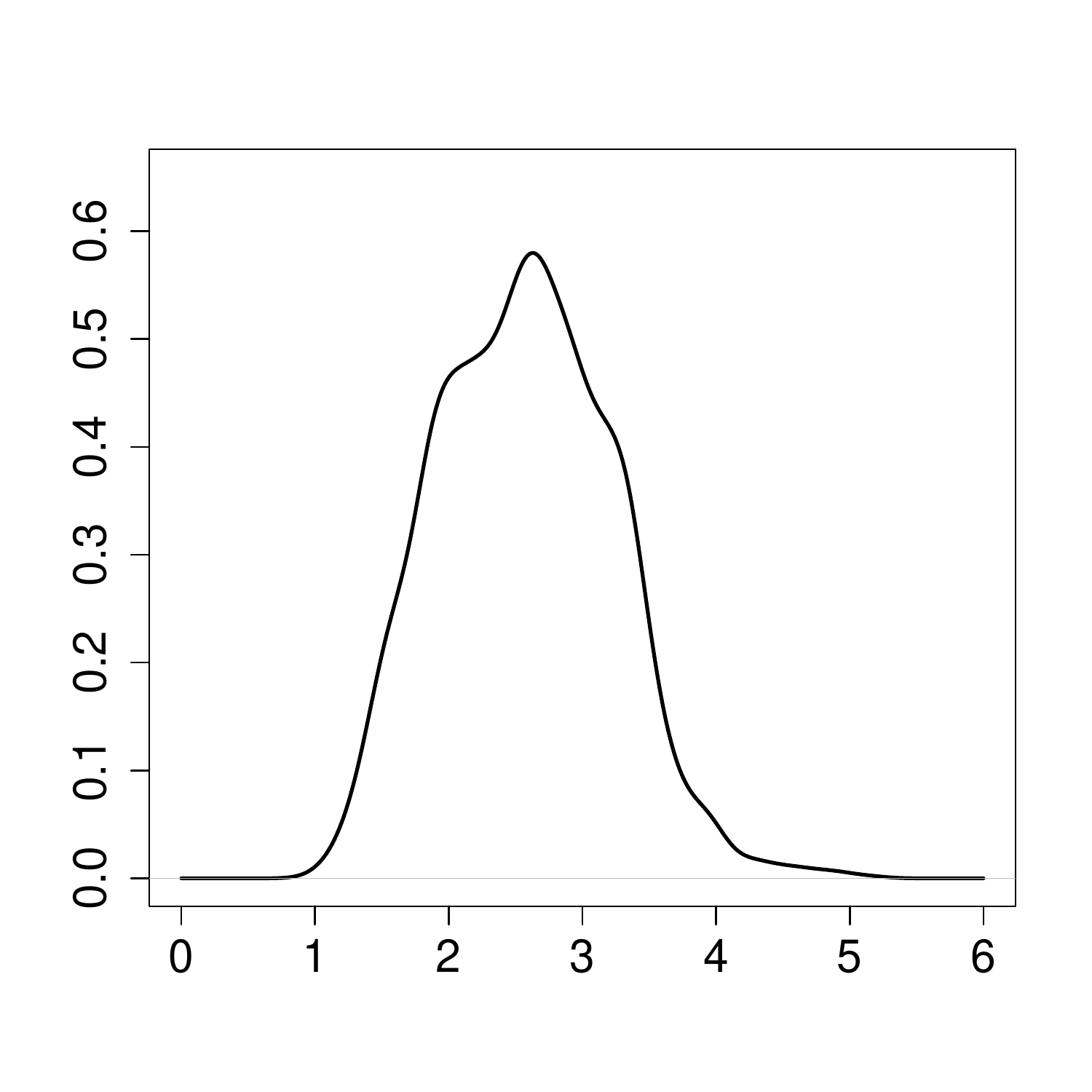}
\includegraphics[width=0.4\linewidth]{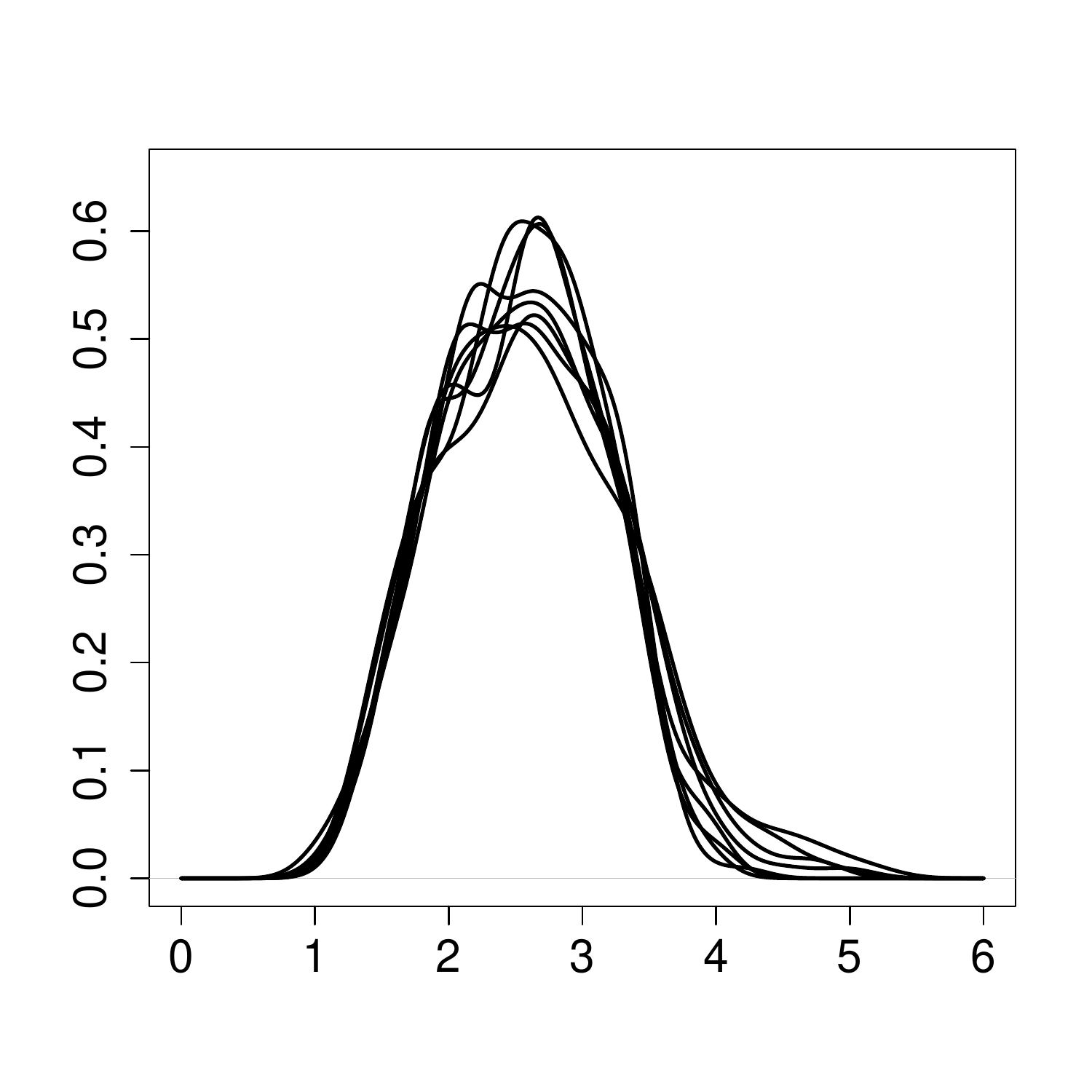}

\caption{A kernel density estimate based on the radii $\rr_n$ (left panel) and eight kernel density estimates based on the radii associated to the points in the eight sets for the subdivision of $W$ obtained by dividing its sides into half parts (right panel). } 
\label{fig:ap:radii}
\end{figure}

The key assumption of homogeneity of $\x_n$ and $\rr_n$ conditioned on $\x_n$ is investigated in this section.
Figure \ref{fig:ap:generators} depicts $\x_n$ together with the observation window and the projections of $\x_n$ onto $xy$, $xz$ and $yz$ planes. The figure indicates spatial homogeneity of $\x_n$. In Figure \ref{fig:ap:radii}, the left panel shows a kernel density estimate of the radii distribution and, for comparison purposes, the right panel shows eight kernel density estimates of the radii distributions corresponding to a subdivision of $W$ into the $2^3=8$ subsets obtained by dividing the three sides of the rectangular region $W$ into half sides. For the number of points per subset, the mean is $232$, the minimum is $219$ and the maximum is $270$. The similarity of the eight density estimates is in accordance with the assumption 
that the distribution of $\rr_n$ conditioned on $\x_n$ is homogeneous.

\section{Strauss process}\label{ap:3}

\begin{figure}[h!]
\center
\includegraphics[width=0.3\linewidth]{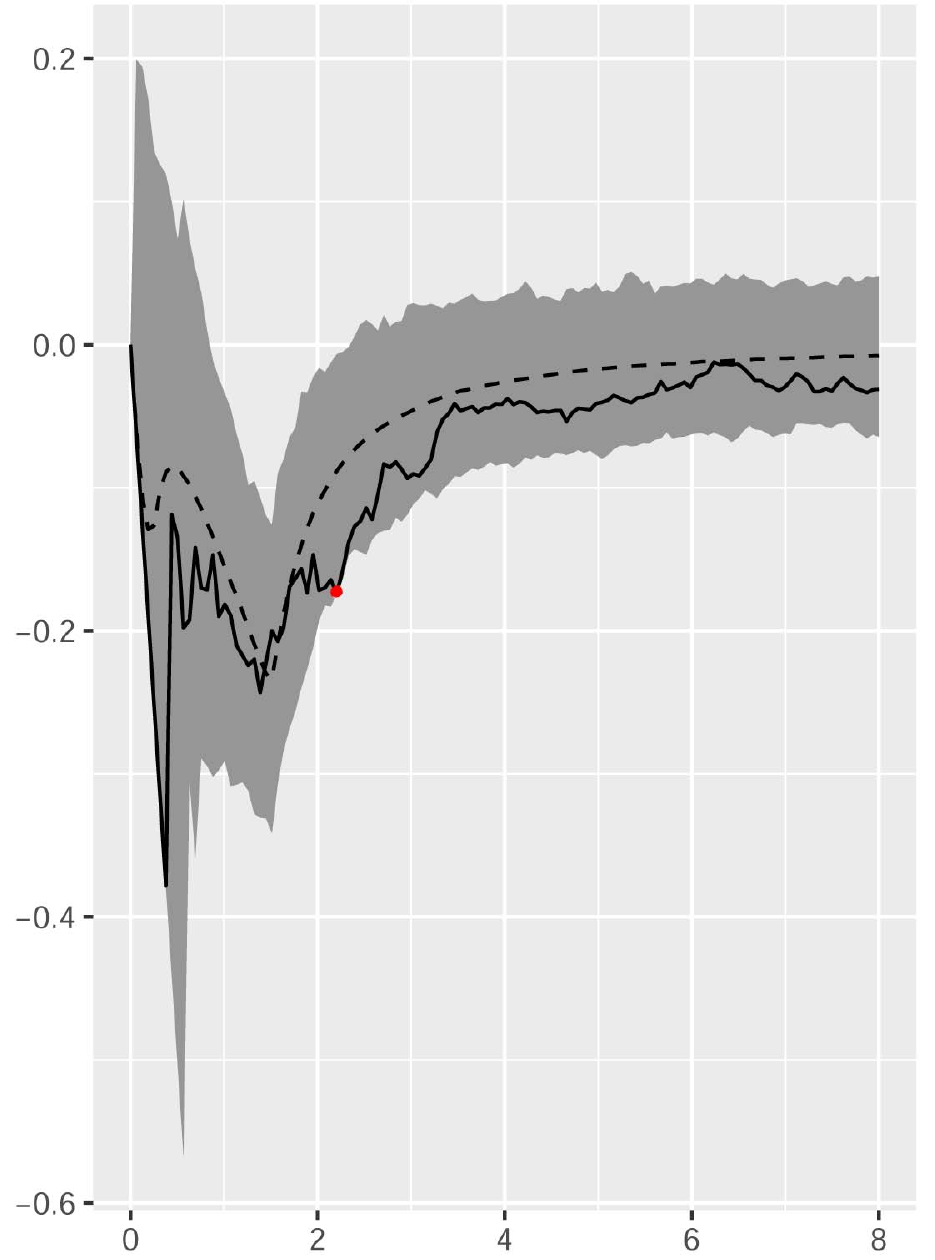}
\includegraphics[width=0.3\linewidth]{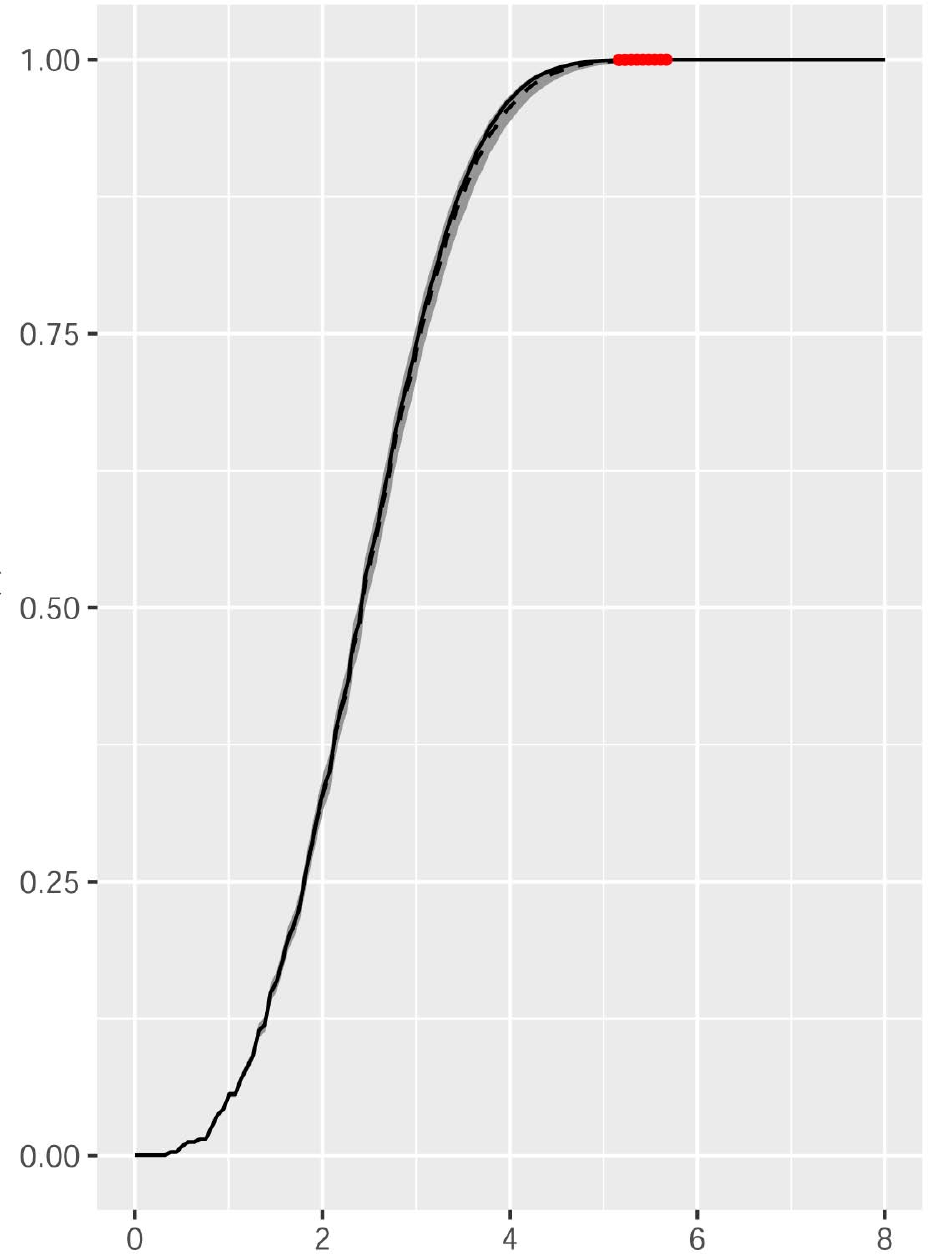}
\includegraphics[width=0.3\linewidth]{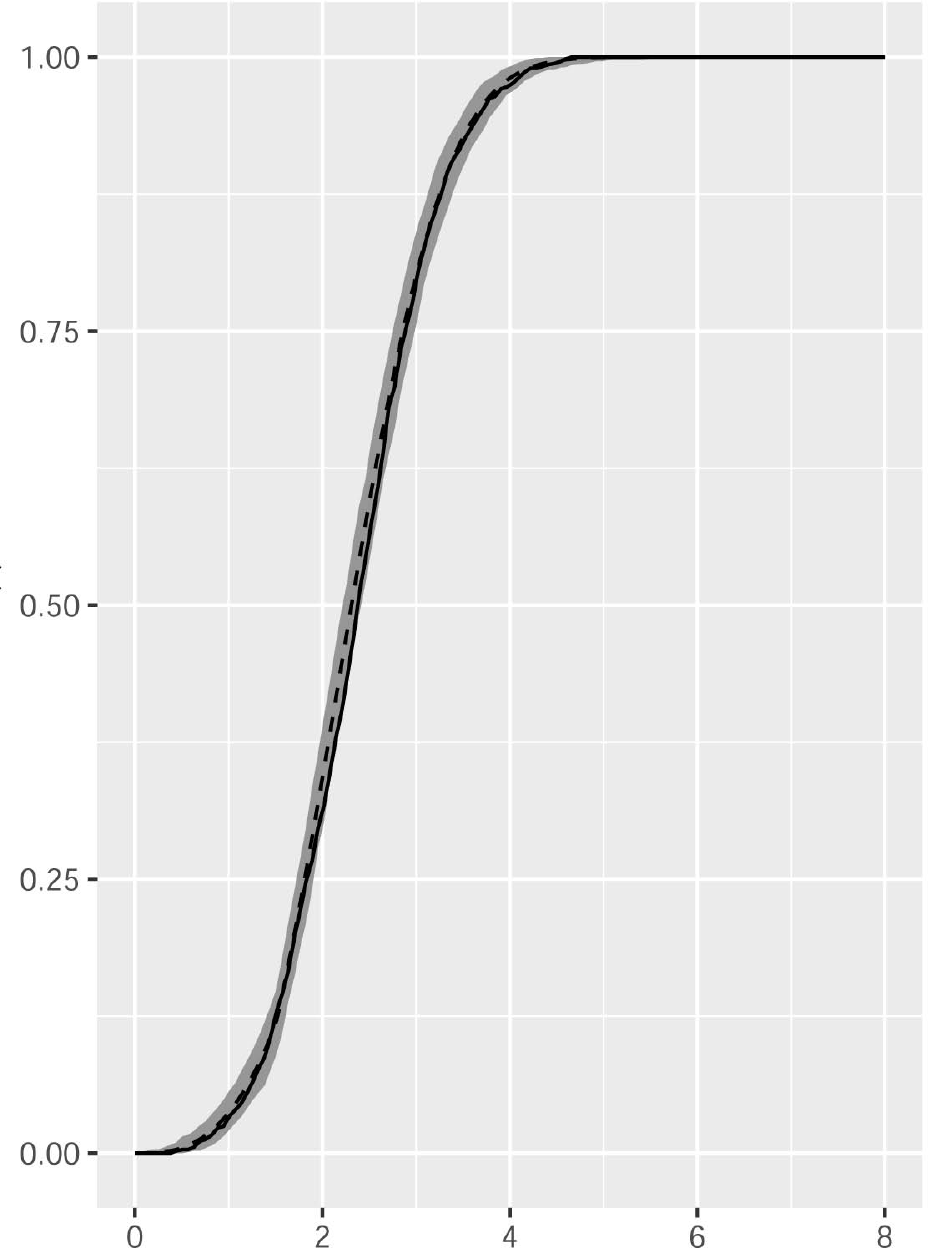}
\caption{From left to right, empirical functional summary statistics $\hat L(t)-t$, $\hat F(t)$ and $\hat G(t)$ (solid lines) and simulated $95\%$-global envelope (grey region) obtained under a fitted Strauss process. Dashed lines are averages of the simulated functional summary statistics.  } 
\label{fig:ap:PP2_GET}
\end{figure}

Figure \ref{fig:ap:PP2_GET} shows three empirical functional summary statistics together with $95\%$-global envelopes obtained by simulations under the Strauss process model $\mathcal M_2$ (the multiscale process with $d=2$) fitted by maximum pseudolikelihood.  The corresponding $p$-value obtained by the global area rank envelope test is $4.4\%$. 


\section{Pseudolikelihood functions}\label{ap:4}

\subsection{Pseudolikelihood for the points}
As in the paper, consider a point process with a given realization $\y_m$ in the observation window $W$ and modelled by a parametric density $p_\vartheta$ (with respect to the unit rate Poisson process) where $\vartheta$ is an unknown real parameter vector. Assume $p_\vartheta$ is hereditary, i.e., for $u\in W\setminus\y_m$ we have $p(\{u\} \cup \y_m)>0$ whenever $p(\y_m)>0$. Then the pseudolikelihood function for the points is
$${\cal{PL}}(\vartheta) = \exp{\left(|W| - \int_W \lambda^{\star}_{\vartheta}(u,\y_m)\,\mathrm du\right)} \prod_{y_j \in \y_m} \lambda^{\star}_{\vartheta}(y_j,\y_m \setminus y_j)$$
where $\lambda^{\star}_\vartheta(u,\y_m) = {p_\vartheta(u \cup \y_m)}/{p_\vartheta(\y_m)}$ is the Papangelou conditional intensity (setting $0/0=0$). Denoting the natural logarithm by $\log$, the log pseudolikelihood function in the case of the multiscale process ${\cal M}_d$ (see \eqref{PIPP} with $d>1$) becomes
\begin{equation*}
\begin{split}
\log {\cal{PL}}(\beta,\gamma) = & |W| - \int_W \beta \prod_{i=1}^{d-1} \gamma_i^{t_{\delta_i}(u,\y_m) - t_{\delta_{i-1}}(u,\y_m)}\,\mathrm du \\ & + m \log \beta + 2\sum_{i=1}^{d-1} \left(S_{\delta_i}(\y_m) - S_{\delta_{i-1}}(\y_m) \right)\log \gamma_i ,\\
\end{split}
\end{equation*}
where $t_{\delta}(u,\y_m) = \sum_{y_j \in \y_m} \mathbb{I}_{\{0 < \|u,y_j\|\leq \delta\}}$ and $S_{\delta}(\y_m) = \sum_{u \in \y_m} t_{\delta}(u,\y_m)/2$.

\subsection{Pseudolikelihood for the radii given the points}

Denoting the conditional density \eqref{MR} by $p_\theta$, the pseudolikelihood function for the radii $\ttt_m$ given the points $\y_m$ is
$${\cal{PL}}(\theta)= \prod_{j=1}^n p_{\theta}(t_j \mid y_j, (y_k,t_k) \text{ with } k \neq j) = \prod_{j=1}^n \frac{h_{\theta}(\ttt_m \mid \y_m)}{\int_0^{6} h_{\theta}(\ttt_m \mid \y_m)\,\mathrm dt_j}$$
where $p_\theta(y_m,\ttt_m) \propto h_{\theta}(\ttt_m \mid \y_m)$.
Assume that $(\y_m,\ttt_m)$ is feasible, i.e., all Laguerre cells $C(y_j,t_j\mid \y_m^*,\ttt_m^*)$ are nonempty; we denote this property by $(\y_m,\ttt_m) \notin {\cal{L}}(\emptyset)$. Then the log pseudolikelihood function for the radii distribution given the points is 
\begin{equation*}
\begin{split}
\log {\cal{PL}}(\theta) = & n \sum_{i=1}^q \theta_i H_i(\y_m,\ttt_m) \\ & - \sum_{j=1}^m \log \int_0^{6} \mathbb{I}_{\{ (\y_m,\ttt_m^{j,u}) \notin {\cal{L}}(\emptyset) \} } \exp\left( \sum_{i=1}^q \theta_i H_i(\y_m,\ttt_m^{j,u})\right)\,\mathrm du \\
\end{split}
\end{equation*}
where $\ttt_m^{j,u} = (t_1,\ldots,t_{j-1},u,t_{j+1},\ldots,t_m)$.

\section{A model for the radii given the points where the `beta' term does not appear}\label{ap:5}


In order to see the effect of omitting the `beta' term in the density \eqref{MR} for the radii conditioned on the points, we fitted the model `nof$+$surf$+$dvol' corresponding to $H=(\sum{\mathrm{nof}},\sum{\mathrm{surf}},\sum{\mathrm{dvol}})$ in \eqref{MR}. Since this is a three-dimensional model, we compare with the results for the fitted three-dimensional model `beta+dvol' in the paper.

\begin{figure}[h!]
\includegraphics[width=1\linewidth]{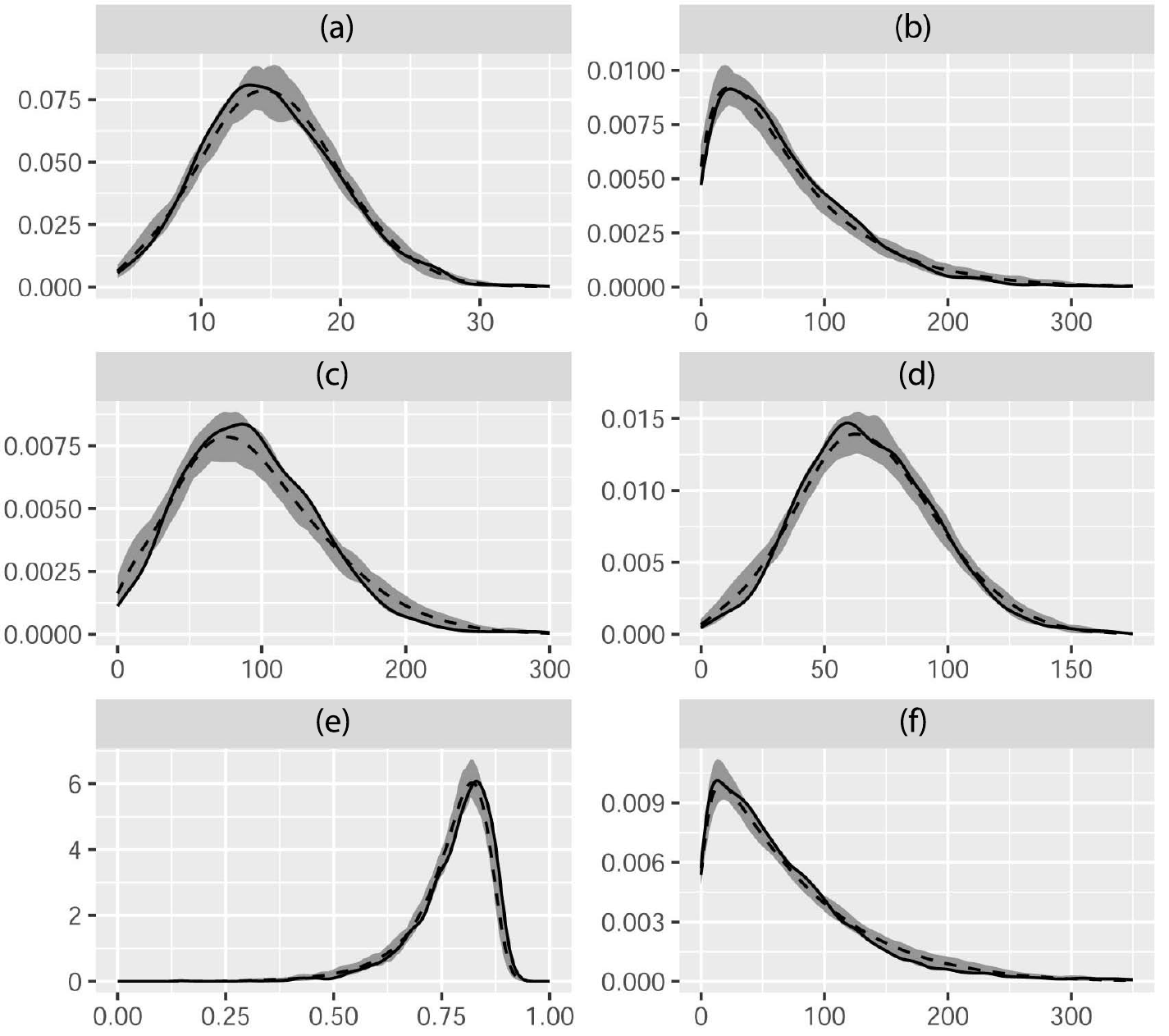}
\caption{
Estimated densities of tessellation characteristics, namely nof (a), vol (b), surf (c), tel (d), spher (e) and dvol (f). The solid lines are the functions based on the data $(\x_n,\rr_n)$ and the grey regions are 95\%-global envelopes under the fitted model `nof+surf+dvol'. Dashed lines are averages of the simulated densities.
} 
\label{fig:ap:rad2_GET}
\end{figure}

Figure~\ref{fig:ap:rad2_GET} shows empirical 
kernel estimates of the densities for the six characteristics in $\mathcal L$ (see \eqref{list}) together with a  
$95\%$-global envelope obtained when concatenating all six empirical densities and  using the \texttt{R}-package \texttt{GET} 
with $499$ simulations of the fitted joint model for $\x_n$ and $\rr_n$. The plots show that the functions based on the data
are completely covered by the envelope. Moreover, the corresponding $p$-value obtained by the global area rank envelope test 
is 7.7\%, which is less than the $p$-value 9.8\% obtained for the fitted `beta+dvol' model. 
Finally, the maximum pseudolikehood estimates are $\hat{\theta}_1 = 3.209\times 10^{-2}$, $\hat{\theta}_2 = -2.993\times 10^{-2}$ and $\hat{\theta}_3 = 2.049\times 10^{-4}$, and the value of the maximized log pseudolikelihood function is $-2470.12$, which is smaller than the corresponding value of $-2468.76$ for the fitted `beta+dvol' model. 

\bibliography{mybibfile}

\end{document}